\documentclass[
  journal=pasa,
  manuscript=research-paper, %% or "review"
  year=2020,
  volume=37,
]{cup-journal}

\usepackage{microtype,siunitx,booktabs}
\def\arcsec{\hbox{$^{\prime\prime}$}}
\def\arcmin{\hbox{$^{\prime}$}}
\newcommand\farcs{\hbox{$.\!\!^{\prime\prime}$}}
\newcommand{\gudel}{G\"udel}
\newcommand{\fracpol}{$|S_V|/S_I$}
\DeclareSIUnit\percent{per~cent}
\newcommand{\stab}{\texttt{stars} table}
\newcommand{\rtab}{\texttt{radio} table}

\newcommand{\srsctot}{839}
\newcommand{\srscdet}{3,405}
\newcommand{\srsccm}{727}
\newcommand{\srscdist}{800}

\newcommand{\srscero}{530}
\newcommand{\srscsingles}{398}
\newcommand{\srscmultis}{441}

\usepackage{hyperref}
\hypersetup{colorlinks,citecolor=blue,linkcolor=blue,urlcolor=blue}

\usepackage{array}
\newcolumntype{P}[1]{>{\raggedright\arraybackslash}p{#1}}
\newcolumntype{A}[1]{>{\raggedleft\arraybackslash}p{#1}}

\usepackage{amssymb}

\usepackage{caption}
\usepackage{subcaption}
\DeclareCaptionFormat{cont}{#1 (cont.)#2#3\par}
\usepackage{multirow}
\usepackage{pdflscape}
\usepackage{longtable}

\title{The Sydney Radio Star Catalogue: properties of radio stars at megahertz to gigahertz frequencies}

\author{Laura~N. Driessen}
\affiliation{Sydney Institute for Astronomy, School of Physics, The University of Sydney, New South Wales 2006, Australia}
\email[L.~N. Driessen]{Laura.Driessen@sydney.edu.au}

\author{Joshua Pritchard}
\affiliation{Sydney Institute for Astronomy, School of Physics, The University of Sydney, New South Wales 2006, Australia}

\author{Tara Murphy}
\affiliation{Sydney Institute for Astronomy, School of Physics, The University of Sydney, New South Wales 2006, Australia}

\author{George Heald}
\affiliation{Australian Telescope National Facility, CSIRO, Space and Astronomy, PO Box 1130, Bentley, WA 6102, Australia.}

\author{Jan Robrade}
\affiliation{Hamburger Sternwarte, Universität Hamburg, Gojenbergsweg 112, 21029 Hamburg, Germany}

\author{Barnali Das}
\affiliation{Australian Telescope National Facility, CSIRO, Space and Astronomy, PO Box 1130, Bentley, WA 6102, Australia.}

\author{Stefan Duchesne}
\affiliation{Australian Telescope National Facility, CSIRO, Space and Astronomy, PO Box 1130, Bentley, WA 6102, Australia.}

\author{David~L.~Kaplan}
\affiliation{Center for Gravitation, Cosmology, and Astrophysics, Department of Physics, University of Wisconsin-Milwaukee, P.O. Box 413, Milwaukee, WI 53201, USA}

\author{Emil Lenc}
\affiliation{Australian Telescope National Facility, CSIRO Astronomy and Space Science, PO Box 76, Epping, NSW 1710, Australia}

\author{Christene R. Lynch}
\affiliation{Department of Physics and Astronomy, University of North Carolina Asheville, Asheville, NC 28804, USA}

\author{Benjamin J. S. Pope}
\affiliation{School of Mathematics and Physics, University of Queensland, St Lucia, QLD~4072, Australia}
\alsoaffiliation{Centre for Astrophysics, University of Southern Queensland, West Street, Toowoomba, QLD~4350, Australia}

\author{Kovi Rose}
\affiliation{Sydney Institute for Astronomy, School of Physics, The University of Sydney, New South Wales 2006, Australia}
\alsoaffiliation{Australian Telescope National Facility, CSIRO Astronomy and Space Science, PO Box 76, Epping, NSW 1710, Australia}

\author{Beate Stelzer}
\affiliation{Institut für Astronomie und Astrophysik, Eberhard Karls Universität Tübingen, Sand 1, 72076 Tübingen, Germany}

\author{Yuanming Wang}
\affiliation{Centre for Astrophysics and Supercomputing, Swinburne University of
Technology, Hawthorn, Victoria, 3122, Australia}

\author{Andrew Zic}
\affiliation{Australian Telescope National Facility, CSIRO Astronomy and Space Science, PO Box 76, Epping, NSW 1710, Australia}

\doi{XX.XXXX/pasa.XXXX.XX}

\received {dd Mmm YYYY}
\revised  {dd Mmm YYYY}
\accepted {dd Mmm YYYY}
\published{dd Mmm YYYY}


\begin{document}

\begin{abstract}
We present the Sydney Radio Star Catalogue, a new catalogue of stars detected at megahertz to gigahertz radio frequencies. It consists of \srsctot\ unique stars with \srscdet\ radio detections, more than doubling the previously known number of radio stars. We have included stars from large area searches for radio stars found using circular polarisation searches, cross-matching, variability searches, and proper motion searches as well as presenting hundreds of newly detected stars from our search of Australian SKA Pathfinder observations. The focus of this first version of the catalogue is on objects detected in surveys using SKA precursor instruments; however we will expand this scope in future versions. The \srsctot\ objects in the Sydney Radio Star Catalogue are distributed across the whole sky and range from ultracool dwarfs to Wolf-Rayet stars. We find that the radio luminosities of cool dwarfs are lower than the radio luminosities of more evolved sub-giant and giant stars. We use X-ray detections of \srscero\ radio stars by the eROSITA soft X-ray instrument onboard the SRG spacecraft to show that almost all of the radio stars in the catalogue are over-luminous in the radio, indicating that the majority of stars at these radio frequencies are coherent radio emitters. The Sydney Radio Star Catalogue can be found in Vizier or at \href{www.radiostars.org}{https://radiostars.org}.
\end{abstract}

\section{Introduction}
\label{sec: intro}

Radio emission has been detected from stars across the main sequence and beyond \citep[e.g.][]{2019PASP..131a6001M}, from ultracool dwarfs \citep[e.g.][]{2001Natur.410..338B,2012ApJ...747L..22R,2015ApJ...808..189W,2018ApJS..237...25K} to Wolf-Rayet stars \citep[e.g.][]{1976ApJ...203L..35S,1986ApJ...303..239A,2005ApJ...623..447D}. Radio emission provides information about the coronae and magnetic fields of stars \citep[][]{1985ARA&A..23..169D} and has the potential to reveal magnetic interaction between stars and planets \citep[e.g.][]{2020NatAs...4..577V,2024MNRAS.528.2136S}. Radio stars are a useful tool for linking the optical and radio coordinate reference frames \citep[e.g.][]{1997A&AS..122..529W}, and are likely to be one of the most common variable objects detected by the Square Kilometre Array after Active Galactic Nuclei (AGN). It is therefore important to identify and study a larger population of radio stars.

The first stars to be detected at radio frequencies were identified in 1963 \citep{1963Natur.198,1963Natur.199..991S}, four years before the discovery of the first pulsar \citep{1968Natur.217..709H}. Despite this, the number of known radio stars has remained low, with one of the key challenges in identifying them being chance coincidence between stars and radio galaxies \citep{2019RNAAS...3...37C}.
The canonical catalogue for radio stars is the Catalogue of Radio Stars \citep[CRS;][]{Wendker_1978,Wendker_1987,Wendker_1995}, which was last updated on 2001 March 26.
It contains 228 radio stars detected at $<3$\,GHz.
The CRS has been an excellent resource; however, recent wide-area radio surveys at $<3$\,GHz, and new radio star identifications, mean that it no longer reflects the population of known radio stars, and a comprehensive update is needed.

Recently, large sky area surveys with the Australian SKA Pathfinder \citep[ASKAP\footnote{\href{https://www.atnf.csiro.au/projects/askap/index.html}{https://www.atnf.csiro.au/projects/askap/index.html}};][]{2021PASA...38....9H} and the Low Frequency Array \citep[LOFAR;][]{2013A&A...556A...2V} have significantly increased the number of known radio emitting stars by using circular polarisation (Stokes \textit{V}), variability and proper motion searches. One reason for this success is that high circular polarisation fraction ($\gtrsim10\%$) can be used to exclude AGN, the dominant type of radio point source by far, since radio-bright AGN have a circular polarisation fraction $\lesssim1\%$  \citep[e.g.][]{1988ARA&A..26...93S,2013MNRAS.435..311O}. Using this method with the LOFAR telescope, \citet{2021NatAs...5.1233C} and \citet{2021A&A...654A..21T} identified 18 previously unknown radio-active M stars, and detected one known radio-active M star and 14 known radio-active RS Canum Venaticorum (RS CVn) binaries. The M dwarfs reported by \citet{2021NatAs...5.1233C} and RS CVns reported by \citet{2021A&A...654A..21T} were included in the LOFAR Two-metre Sky Survey \citep[LoTSS;][]{2017A&A...598A.104S,2022A&A...659A...1S} in Stokes \textit{V} \citep[V-LoTSS][]{vlotss} paper as well as some new radio star identifications for a total of 37 radio identified stars. \citet{2021MNRAS.502.5438P} used ASKAP Stokes \textit{V} detections to identify 23 previously unknown radio stars and 10 known ones.
\citet{2024MNRAS.tmp..161P} expanded on this work and found 36 radio stars by performing multi-epoch Stokes \textit{V} searches using ASKAP.

Untargeted searches for variable radio sources have resulted in detections of flares from eight stars using ASKAP \citep{Yuanming_ShortTimescale} and three stars using the (more) Karoo Array Telescope \citep[MeerKAT;][]{2018ApJ...856..180C}. \citet{Driessen_ProperMotion} identified eight radio stars, two of which had not previously been identified, by searching for changes in radio position due to proper motion.

Cross-matching directly between radio point source surveys and optical sky surveys such as \textit{Gaia} \citep{GaiaMission,GaiaDR3,GaiaDR3_Validation} results in a high chance coincidence rate \citep[][]{2019RNAAS...3...37C} between radio AGN and optical stars. However, the probability of chance coincidence decreases when limited samples are used, such as volume-limited samples. 
For example, cross-matching between the Karl G. Jansky Very Large Array \citep[VLA;][]{perley2011} Faint Images of the Radio Sky at Twenty-centimetres \citep[FIRST;][]{1994ASPC...61..165B,1995ApJ...450..559B} survey and various optical catalogues resulted in the detection of 26 radio stars \citep{1999AJ....117.1568H}. They calculated that 0.96$\pm$0.04 of their matches with  Hipparcos  \citep{hipparcos} were by chance.
At the time, this doubled the number of known radio stars in the area of sky covered by the search. This was a cross-match using a comparatively limited sample as FIRST contains 946,000 sources over 10,000 square degrees ($\sim$90 sources per square degree) while the Hipparcos catalogue contains 118,218 sources over the whole sky (3 sources per square degree). This is compared to LoTSS, for example, which contains 4,396,228 radio sources over 5,635 square degrees (780 sources per square degree) and the \textit{Gaia} DR3 catalogue which contains 1.46 billion sources over the whole sky ($\sim$35,000 sources per square degree).
\citet{2023arXiv231207162Y} successfully cross-matched the radio surveys LoTSS and the VLA Sky Survey \citep[VLASS;][]{2020PASP..132c5001L,2021ApJS..255...30G} with the \textit{Gaia} catalogue of Nearby Stars \citep[GCNS;][]{2021A&A...649A...6G}. {They found 25 LoTSS-GCNS radio stars with an estimated number of chance coincidence matches of $\sim1$ and 65 VLASS-GCNS radio stars with an estimated number of chance coincidence matches of $\sim0.5$}. This shows that we can improve reliability by either cross-matching to a volume-limited stellar sample or by cross-matching the radio source catalogue to a subset of stars.

A catalogue of radio stars will be key to investigating the correlation between X-ray and radio emission at different radio frequencies and for different stellar types. This is because the X-ray emission from active stars reveals information about the coronae and the radio emission reveals information about the magnetic fields. The \gudel--Benz relation is a known correlation between the $\sim$5 - 9 GHz radio luminosity and $\sim$0.1 - 2.4 keV (\textit{ROSAT} band) X-ray luminosity of active stars. The original relation was given by $L_X/L_R \approx 10^{15.5}$ where $L_R$ and $L_X$ are the quiescent radio specific and X-ray luminosities respectively \citep{1993ApJ...405L..63G}. A more recent version is now more commonly used, where $\log_{10}\left[ L_R \right] = 1.36\left( \log_{10}\left[ L_X \right] - 18.96 \right)$ \citep{2014ApJ...785....9W}. The \gudel--Benz relation implies that the mechanism that accelerates non-thermal electrons and the mechanism that heats the corona are the same.

The \gudel--Benz relation has been demonstrated to apply to incoherent, gyrosynchtrotron emission from  RS~CVns, single active cool dwarfs, BY Draconis binaries, and solar flares. At lower frequencies than 5\,GHz, it has been demonstrated that the \gudel--Benz relation no longer applies for some stars \citep{2021NatAs...5.1233C} because the radio emission arises from coherent mechanisms, as coherent radio emission is more luminous and does not contribute to heating the corona. However, some coherently-emitting radio stars do still follow the relation closely \citep{2022ApJ...926L..30V}. More investigation is needed to determine the frequency at which the radio emission mechanism changes and whether there is a different relation at different frequencies. Or it may be that at lower radio frequencies some types of radio star do follow the relation while others do not, regardless of whether the emission is coherent or incoherent. A catalogue of radio stars will be important for this, but so will using radio luminosities from a single instrument and X-ray luminosities from a single instrument. ASKAP has a wide field-of-view and has performed multiple all-sky surveys making it a useful instrument for this task. It is also key to use consistent X-ray measurements with as many stars included as possible. The eROSITA \citep[extended ROentgen Survey with an Imaging Telescope Array,][]{2021A&A...647A...1P} soft X-ray instrument onboard the SRG spacecraft is the ideal instrument for this and the all-sky survey it performs is essential for investigating active stars now and in the future.

In this paper we present a new catalogue of \srsctot\ unique radio stars: the Sydney Radio Star Catalogue (SRSC). This includes hundreds of new radio star identifications found by performing circular polarisation and cross-match searches using ASKAP. We focus on producing a reliable catalogue using only stars from large-scale, wide-field searches for radio stars using SKA precursor instruments. We aim to provide a useful resource for the radio star community, and we plan to update the web version of the catalogue to include published radio stars and future radio star identifications. In Section\,\ref{sec: methods overview} we present the catalogues and search methods used to identify the radio stars included in the SRSC. In Section\,\ref{sec: catalogue format} we present the format of the catalogue. In Section\,\ref{sec: catalogue properties} we present the content and properties of the SRSC. We summarise in Section\,\ref{sec: summary}.

\section{Radio star identification methods}
\label{sec: methods overview}

The catalogue is constructed from samples of confirmed radio stars already published in the literature, as well as newly detected radio stars from the work presented here. We have only included stars from searches for radio stars using SKA precursors for which the detection of radio emission is confirmed and reliable. We have not included candidate radio stars, for which the association is not confirmed. 

In this section we will present the method for and results of searching for new radio star identifications using ASKAP via circular polarisation searches (Section\,\ref{sec: ASKAP circ pol search}) and cross-matching (Section\,\ref{sec: xmatch}). We will also detail what other samples of stars are included in the SRSC (Section\,\ref{sec: other data}) and those that we have chosen not to include for this version of the SRSC (Section\,\ref{sec: data not included}).

\subsection{ASKAP circular polarisation searches}
\label{sec: ASKAP circ pol search}

We include 235 stars discovered via circular polarisation searches of 
Rapid ASKAP Continuum Survey (RACS) data and Variables and Slow Transients with ASKAP \citep[VAST;][]{VAST} Pilot (VASTP)  survey data.
RACS is the ASKAP all-sky radio survey and has been conducted at three frequencies: 887.5\,MHz \citep[RACS-low;][]{2020PASA...37...48M,2021PASA...38...58H}, 1367.5\,MHz \citep[RACS-mid;][]{2023PASA...40...34D}, and at 1655.5\,MHz (RACS-high; Duchesne et al. in prep). RACS-low covers the southern sky up to $+40^{\circ}$ declination and both RACS-mid and RACS-high cover the southern sky up to $+50^{\circ}$ declination. RACS-low and RACS-mid are currently available in the CSIRO ASKAP Science Data Archive (CASDA)\footnote{\href{https://research.csiro.au/casda/}{https://research.csiro.au/casda/}}, we expect that the RACS-high data will be available mid-2024. The initial RACS-high release will be further processed and combined into an individual catalogue as part of a future publication (Duchesne et al. in prep).

The VAST team is performing untargeted searches for variable and transient radio sources in 12-minute, 887.5\,MHz ASKAP observations.
VASTP-low is the low band of the VAST pilot survey at $\SI{887.5}{\mega\hertz}$ which covered a $\SI{5131}{\deg^2}$ footprint 
with a varying degree of coverage over 15 epochs (see \citet{Murphy2021} and \citet{2024MNRAS.tmp..161P} for details).
The circular polarisation search presented here used RACS-low and RACS-mid data as well as the VASTP data.
We refer the reader to
\citet{2021MNRAS.502.5438P} and \citet{2024MNRAS.tmp..161P} for a detailed
discussion of the circular polarisation search method, artefact
rejection, and classification procedure, and summarise the main
points here. In each search the {\sc selavy} \citep{2012MNRAS.421.3242W,2012PASA...29..371W} source
finder package was used to extract lists of 2D Gaussian source components
from the Stokes $I$ and $V$ images, and sources
with significant circular polarisation were identified by crossmatching the
extracted Stokes $V$ components with their counterparts in Stokes
$I$. The candidates were restricted to Stokes $I$-$V$
associations that were greater than 5$\sigma$ in both Stokes $I$
and $V$ where $\sigma$ is the local RMS noise, and with
fractional circular polarisation \fracpol\ > \SI{6}{\percent},
corresponding to 10 times the median polarisation leakage of
Stokes $I$ into $V$ \citep{2024MNRAS.tmp..161P}. 

After manual rejection of Stokes $I$-$V$ associations with artificially high \fracpol\ 
caused by imaging artefacts and noise, 48 radio stars were identified in RACS-low, 
65 radio stars in RACS-mid, and 184 detections of 36 stars over the multi-epoch 
VASTP-low survey.

\subsection{ASKAP catalogue cross-matching}
\label{sec: xmatch}

We used cross-matching between ASKAP radio point source positions and selected star catalogues to search for radio stars. We reduced the chance-coincidence probability of the cross-matches by using star catalogues that were either volume limited, or only included a subset of stars. This approach resulted in the identification of 734 stars. In this section we describe this method,  the ASKAP and star catalogues we used for cross-matching, and our results.

\subsubsection{ASKAP sample preparation}
\label{sec: xmatch ASKAP cats}

We used all publicly available ASKAP data in CASDA as of 2023 October 16, as well as RACS-low, RACS-mid, and RACS-high data.
The ASKAP data available in CASDA varies in integration time and frequency. The integration times range from just under a minute to thirteen hours. The observations have frequency bands with centre frequencies of between 700 and 1700\,MHz MHz. We produced combined ASKAP source catalogues from the publicly available {\sc selavy} \citep{2012MNRAS.421.3242W,2012PASA...29..371W} catalogues. Each combined catalogue contained observations with integration times from $0-1$ hours, $1-2$ hours and so on. This is because source density increases with integration time, due to improved sensitivity. As a result, the chance coincidence probability is higher in longer integration observations, and means that a smaller match radius is needed. We removed duplicate sources from each sample by removing sources where the separation was less than the average of the semi-major axis of the two sources. We kept only one instance of each source to prevent multiple cross-matches to a single star, as this would affect our reliability statistics. Table\,\ref{tab: ASKAP catalogue lengths} shows the total number of sources in each catalogue.

\begin{table}
\caption{Number of sources, $N_\mathrm{ASKAP}$, in each ASKAP samples. $N_\mathrm{ASKAP}$ is the total number of ASKAP sources in each catalogue, excluding duplicate sources. $N_\mathrm{ASKAP,!G}$ is the number of ASKAP sources in each catalogue after the Milliquas and 6dFGSzDR3 cross-matched sources have been removed. $N_{\mathrm{ASKAP,!G,PS}}$ is the number of ASKAP sources in each catalogue after the Milliquas and 6dFGSzDR3 cross-matched sources have been removed and after removing sources where $S_{\mathrm{int}} / S_{\mathrm{peak}} > 1.5$.
}
    \centering
    \begin{tabular}{P{1.4cm}rrr}
Sample name & $N_\mathrm{ASKAP}$ & $N_\mathrm{ASKAP,!G}$ & $N_\mathrm{ASKAP,!G,PS}$ \\
\hline
\hline
RACS-LOW & 2,665,933 & 2,617,851 & 1,733,301 \\
RACS-MID & 3,107,143 & 3,038,033 & 2,118,761 \\
RACS-HIGH & 3,204,704 & 3,131,871 & 2,455,364 \\
00-01hour & 2,201,419 & 2,170,113 & 2,169,828 \\
01-02hour & 119,463 & 117,924 & 117,907 \\
02-03hour & 1,200,003 & 1,180,784 & 1,180,561 \\
03-04hour & 478,887 & 467,810 & 467,762 \\
04-05hour & 137,416 & 134,691 & 134,681 \\
05-06hour & 573,419 & 563,380 & 563,211 \\
06-07hour & 1,346,916 & 1,318,819 & 1,318,689 \\
07-08hour & 190,507 & 187,600 & 187,555 \\
08-09hour & 734,965 & 725,826 & 725,477 \\
09-10hour & 229,746 & 225,730 & 225,655 \\
10-11hour & 2,151,104 & 2,142,785 & 2,141,799 \\
11-12hour & 91,946 & 88,258 & 88,244 \\
12-13hour & 105,013 & 103,614 & 103,588 \\
13-14hour & 38,802 & 38,627 & 38,613 \\
    \hline
    \end{tabular}
    \label{tab: ASKAP catalogue lengths}
 \end{table}

To reduce contamination from AGN we removed ASKAP sources that were within 2\arcsec\ of a galaxy in the Milliquas \citep{milliquas} and 6dFGSzDR3 \citep{2005ASPC..329...11J,2009MNRAS.399..683J} catalogues. Milliquas (Million Quasars) is a catalogue of 907,144 quasars and AGN. 6dFGSzDR3 is a catalogue of 125,071 galaxies, including redshifts and velocities for many of them. We also removed resolved ASKAP sources where $S_{\mathrm{int}} / S_{\mathrm{peak}} > 1.5$, where $S_{\mathrm{int}}$ is the integrated flux density and $S_{\mathrm{peak}}$ is the peak flux density.

\subsubsection{Star catalogue preparation}
\label{sec: star cat prep}

The four stellar catalogues that we cross-matched the ASKAP source catalogues to are:
\begin{itemize}
    \item the Fifth Catalogue of Nearby Stars \citep[CNS5;][]{2023A&A...670A..19G};
    \item the stellar content of the \textit{ROSAT} all-sky survey catalogue \citep[][hereafter ``\textit{ROSAT} stellar'']{2022A&A...664A.105F};
    \item the \textit{XMM-Newton} slew survey \citep[][hereafter ``\textit{XMM-Newton} stellar'']{2018A&A...614A.125F};
    \item the Galactic Wolf-Rayet Catalogue\footnote{\href{https://pacrowther.staff.shef.ac.uk/WRcat/}{https://pacrowther.staff.shef.ac.uk/WRcat/}} \citep{2015MNRAS.447.2322R}.
\end{itemize}
The number of sources in each is given in Table\,\ref{tab: Optical catalogue lengths}.

The CNS5 is a catalogue of stars within 25\,pc of the Sun. It incorporates \textit{Gaia} Early Data Release 3 \citep[EDR3;][]{GaiaEDR3_Summary, GaiaEDR3_Validation} data, Hipparcos data \citep{hipparcos} and data from ground-based near-infrared surveys. The CNS5 contains 5\,931 objects: 5\,230 stars (including the Sun) and 701 brown dwarfs. Of the stars, 20 are giant stars, 264 are white dwarfs and 3\,760 are M-dwarfs.
CNS5 is statistically complete down to 19.7 mag in G-band and 11.8 mag in Wide-field Infrared Survey Explorer \citep[\textit{WISE}][]{2010AJ....140.1868W} W1-band absolute magnitudes. The white dwarf sample is statistically complete.
\citet{2023A&A...670A..19G} assume that the expected star-to-brown dwarf ratio is $\sim5$, which means that approximately a third of the brown dwarfs within 25\,pc of the Sun are yet to be discovered.

The \textit{ROSAT} stellar catalogue is a catalogue of X-ray emitting stars identified by cross-matching between the \textit{ROSAT} All-Sky Survey \citep[RASS;][]{ROSAT_RASS} and the \textit{Gaia} EDR3 catalogue.
We use $p_{ij}$ and $p_{\mathrm{stellar}}$, where $p_{ij}$ is the posterior probability that the $j$th counterpart to a \textit{ROSAT} source is the correct identification and
$p_{\mathrm{stellar}}$ is the probability that the counterpart to the \textit{ROSAT} detection is the correct identification, to include only reliable X-ray identifications.
We used only those \textit{ROSAT} stellar sources with $p_{\mathrm{stellar}}>0.51$ and $p_{ij}>0.5$ where the completeness and reliability for the cross-matches are over 93 per cent, resulting in a catalogue containing 28\,109 sources.
We also removed sources flagged as subdwarfs
% as suggested by \citet{2022A&A...664A.105F} 
and kept only a single detection for those sources with multiple X-ray detections, resulting in 27\,881 sources. 

The \textit{XMM-Newton} stellar catalogue contains cross-matches between \textit{XMM-Newton} slew survey sources \citep{2008A&A...480..611S} and stars.
It contains 5,920 sources with a completeness of 96.3 per cent and a reliability of 96.7 per cent.
We filtered this catalogue by including only one-to-one matches between an \textit{XMM-Newton} source and a star.
We excluded those sources that were flagged to have, for example, an accreting object within 30\arcsec, an AGN within 30\arcsec, or other issues identified.
We did this by excluding sources that had a flag in the Stellar Flag (StFlg) column, as suggested by \citet{2008A&A...480..611S}.
This resulted in a catalogue containing 5,042 stars cross-matched to \textit{XMM-Newton} sources.

The Galactic Wolf-Rayet Catalogue contains 669 Wolf-Rayet stars. We included this catalogue as some Wolf-Rayet stars are known to emit at radio wavelengths \citep[][]{2000MNRAS.319.1005D}, such as Apep \citep{2019NatAs...3...82C}, WR 48a \citep{2012MNRAS.421.3418H} and HD~93129A \citep{2015A&A...579A..99B}.

\begin{table}
\caption{Number of sources, $N_{star}$, in the four star catalogues used for cross-matching.}
    \centering
    \begin{tabular}{lrr}
Star catalogue & $N_{star}$ &  $N_{star}$ (Dec $<+50^{\circ}$) \\
\hline
\hline
CNS5 & 5\,908 & 5\,193 \\
\textit{ROSAT} Stellar & 27\,881 & 23\,017 \\
\textit{XMM-Newton} Stellar & 5\,042 & 4\,400 \\
Wolf-Rayet & 669 & 653 \\
    \hline
    \end{tabular}
    \label{tab: Optical catalogue lengths}
 \end{table}

We prepared each of the star catalogues to make sure they were in a consistent format before cross-matching.
For example, some catalogues do not include proper motion information, and others only give a star name, with no further stellar parameters. 
For each catalogue we used the source names provided by the catalogue to identify the SIMBAD \citep{simbad}, \textit{Gaia} DR3 \citep{GaiaMission,GaiaDR3,GaiaDR3_Validation}, DR2 \citep{GaiaDR2} and DR1 \citep{GaiaDR1}, Tycho \citep{TYCHO_2,TYCHO_2_Validation}, Hipparcos \citep{hipparcos}, 2MASS \citep{2MASS}, GCVS \citep{GCVS}, and UCAC4 \citep{UCAC4} names if available. We did this using \texttt{astroquery SIMBAD}\footnote{\href{https://astroquery.readthedocs.io/en/latest/simbad/simbad.html}{https://astroquery.readthedocs.io/en/latest/simbad/simbad.html}} to find object IDs based on the object names. We did not attempt position-based identifications as this risks matching to an incorrect source due to proper motion. If we could not find a name-based match, we used the astrometric information from the original catalogue. For example, if we could not find a name for a CNS5 source we used the astrometric information from the CNS5 catalogue itself.
For each source we provide the SIMBAD, \textit{Gaia} DR3 and DR2, Tycho, Hipparcos, 2MASS, GCVS, and UCAC4 names where available. We also provide the astrometric information from one of \textit{Gaia} DR3, Hipparcos, UCAC4, \textit{Gaia} DR2, or DR1 for each star. We chose the source of the astrometric information based on which catalogue provided astrometric information. For example, if a source has a \textit{Gaia} DR3 position and no proper motion, but has a Hipparcos position that does provide proper motion information, we used the Hipparcos information. If the source has a \textit{Gaia} DR3 position and no proper motion information in any catalogue then we used the \textit{Gaia} DR3 position.

\subsubsection{Cross-match reliability}

We performed Monte Carlo simulations to determine the reliability of the cross-matches and the appropriate cross-match radius to identify radio emission from stars. The cross-match radius is the maximum separation between the centroid of the radio source and the proper motion corrected position of the star.

For the Monte Carlo simulation we used two catalogues, the radio catalogue and the star catalogue. For example, RACS-high as the radio catalogue and the CNS5 as the star catalogue. The simulation was then performed as follows:
\begin{enumerate}
    \item Offset the positions of the stars in a random direction and a random offset magnitude within the range described below. \label{step 1}
    \item Match to the radio catalogue using a cross-match radius of $A\arcsec$. \label{step 2}
    \item Count how many cross-matches there are.  \label{step 3}
    \item Repeat steps \ref{step 1} to \ref{step 3} $M$ times for a range of cross-match radii. \label{step 4}
\end{enumerate}
For step\,\ref{step 1} we offset each star position in the catalogue by taking the square root of a number drawn from a random uniform distribution between
$c^{2}$ and ($c+r$)$^{2}$ where $c$ is the shift and $r$ is the radius. We set $c=1$\arcmin\ and $r=30$\arcsec.
This method means that we are drawing from a random distribution that is uniform over the area of an annulus around each star.
We chose this minimum shift value such that we reduced the number of real matches contaminating our random cross-matches. For step\,\ref{step 2} we used the cumulative number of matches. For example, we counted how many matches there were within 1\arcsec, then 2\arcsec\ and so on. In step\,\ref{step 3} we used a one-to-one matching, so each radio source and each star was only counted once. We matched in radius bins up to 30\arcsec\ with 0\farcs1 bins. In step\,\ref{step 4} we repeated this process for each radio catalogue cross-matched to each star catalogue (shown in Table\,\ref{tab: xmatch}). We repeated it 100,000 times each. We did not take proper motion into account for the Monte Carlo simulations as the density of sources on the sky is more important than the true positions. 

\begin{figure}
\caption{Example cumulative cross-match plot for the cross-match between the CNS5 catalogue and RACS-high. We show the results of the 100,000 iteration Monte Carlo simulation and the true, proper motion corrected cross-match. The match radii for 90, 95 and 98 per cent reliability is also shown here.
}
\includegraphics[width=\columnwidth]{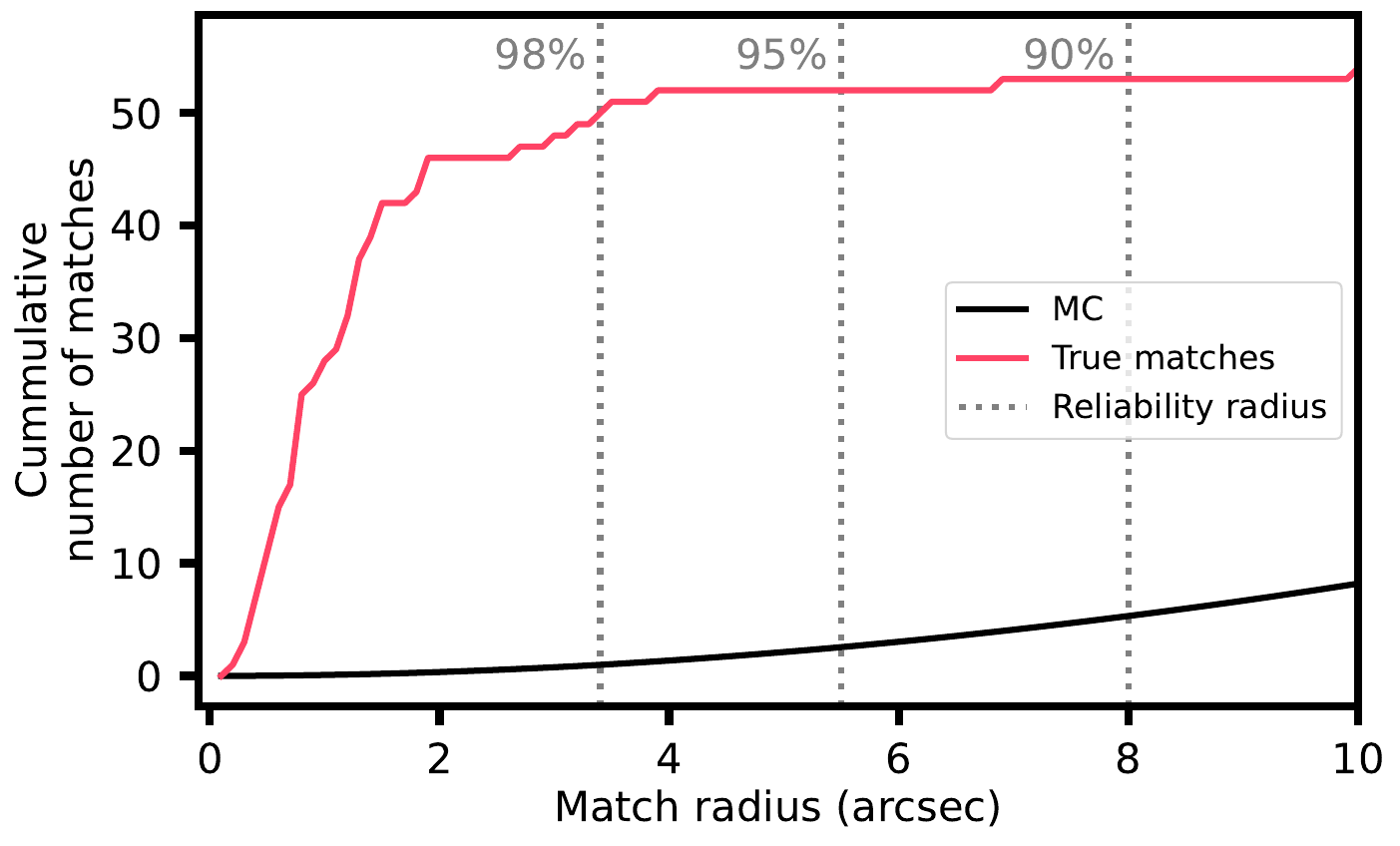}
\label{fig: MC demo}
\end{figure}

We calculated the mean value, $\lambda$, in each radius bin. This gives us the expected number of random matches per radius bin for the cross-match between both complete catalogues. 
This method assumes that none of the matches between the radio catalogue and randomised star catalogue are true matches. 
An example of the cumulative number of matches per bin for a cross-match between RACS-high and CNS5 is shown in Figure\,\ref{fig: MC demo}.

We also performed the binned cross-match between the radio catalogues and the true, proper motion corrected positions of the objects from the star catalogues. We could then calculate the reliability of the cross-matches at different radii up to 30\arcsec. The results of this cross-match between RACS-high and CNS5 are also shown in Figure\,\ref{fig: MC demo}.

The reliability at a given cross-match radius is given by $R=1-N_{\mathrm{random}} / N_{\mathrm{true}}$, where $R$ is the reliability, $N_{\mathrm{true}}$ is the number of matches at a given radius using the true coordinates for both catalogues and $N_{\mathrm{random}}$ is the number of matches at a given radius as a result of the 100,000 iteration Monte Carlo simulation.

We selected the match radius for each radio-star catalogue cross-match, such that the resultant  reliability was 98 per cent. The radius that this corresponds to for each pair of catalogues is shown in Table\,\ref{tab: xmatch}. The typical uncertainty on ASKAP source positions is 2\arcsec. In some cases the 98 per cent reliable match radius is larger than 2\arcsec\ due to the low density of sources in one or both of the catalogues. In these cases we used up to twice the typical ASKAP position uncertainty as the cross-match radius: 4\arcsec. 
For cases where the cross-match radius required for 98 per cent reliability was less than the ASKAP position uncertainty of 2\arcsec\ we did not perform a cross-match.
In some cases there are no matches within 30\arcsec\ of any of the source positions. Those cases have been left blank. The Wolf-Rayet catalogue includes the fewest sources so there are many cases where there are no cross-matches between the Wolf-Rayet source catalogue and the ASKAP source catalogue. This means that the smallest cross-match radius we used was 2\arcsec\ and the largest cross-match radius we used was 4\arcsec. 

\begin{table*}
    \caption{Results of the Monte Carlo analysis and ASKAP-star catalogue cross-matches.
    For each ASKAP-star catalogue cross-match we show the 98\% radius, the cross-match radius used, and the number of matches found.
    }
\centering
\begin{tabular}{p{1.3cm}|A{0.8cm}A{0.8cm}A{1.0cm}|A{0.8cm}A{0.8cm}A{1.0cm}|A{0.8cm}A{0.8cm}A{1.0cm}|A{0.8cm}A{0.8cm}A{1.0cm}}
 & \multicolumn{3}{c|}{CNS5} & \multicolumn{3}{c|}{\textit{ROSAT} stellar} & \multicolumn{3}{c|}{\textit{XMM-Newton} stellar} & \multicolumn{3}{c}{Wolf-Rayet} \\
\cline{2-13}
ASKAP Catalogue & 98\% radius & Match radius & Num. of matches & 98\% radius & Match radius & Num. of matches & 98\% radius & Match radius & Num. of matches & 98\% radius & Match radius & Num. matches \\
\hline
RACS-high & 3.4\arcsec & 3.4\arcsec & 50 & 3.5\arcsec & 3.5\arcsec & 263 & 5.2\arcsec & 4.0\arcsec & 115 & 6.2\arcsec & 4.0\arcsec & 13 \\
RACS-low & 2.3\arcsec & 2.3\arcsec & 16 & 0.8\arcsec & -- & -- & 2.7\arcsec & 2.7\arcsec & 24 & -- & -- & -- \\
RACS-mid & 3.1\arcsec & 3.1\arcsec & 37 & 3.2\arcsec & 3.2\arcsec & 181 & 4.4\arcsec & 4.0\arcsec & 74 & 5.8\arcsec & 4.0\arcsec & 9 \\
00-01hour & 3.3\arcsec & 3.3\arcsec & 44 & 2.7\arcsec & 2.7\arcsec & 125 & 3.6\arcsec & 3.6\arcsec & 54 & 2.0\arcsec & 2.0\arcsec & 6 \\
01-02hour & 5.7\arcsec & 4.0\arcsec & 4 & 3.8\arcsec & 3.8\arcsec & 8 & 6.7\arcsec & 4.0\arcsec & 4 & 3.5\arcsec & 3.5\arcsec & 1 \\
02-03hour & 3.0\arcsec & 3.0\arcsec & 18 & 2.4\arcsec & 2.4\arcsec & 36 & 3.6\arcsec & 3.6\arcsec & 19 & -- & -- & -- \\
03-04hour & 1.1\arcsec & -- & -- & 0.9\arcsec & -- & -- & 2.6\arcsec & 2.6\arcsec & 4 & 8.7\arcsec & 4.0\arcsec & 1 \\
04-05hour & 2.0\arcsec & 2.0\arcsec & 1 & 1.9\arcsec & -- & -- & 2.3\arcsec & 2.3\arcsec & 1 & -- & -- & -- \\
05-06hour & 2.2\arcsec & 2.2\arcsec & 5 & 2.7\arcsec & 2.7\arcsec & 23 & 4.2\arcsec & 4.0\arcsec & 11 & 3.3\arcsec & 3.3\arcsec & 3 \\
06-07hour & 2.7\arcsec & 2.7\arcsec & 17 & 1.9\arcsec & -- & -- & 3.3\arcsec & 3.3\arcsec & 26 & 5.7\arcsec & 4.0\arcsec & 4 \\
07-08hour & 3.7\arcsec & 3.7\arcsec & 4 & 3.7\arcsec & 3.7\arcsec & 11 & 4.5\arcsec & 4.0\arcsec & 5 & -- & -- & -- \\
08-09hour & 2.3\arcsec & 2.3\arcsec & 7 & 3.9\arcsec & 3.9\arcsec & 61 & 5.1\arcsec & 4.0\arcsec & 23 & 24.9\arcsec & 4.0\arcsec & 0 \\
09-10hour & 3.0\arcsec & 3.0\arcsec & 4 & 2.4\arcsec & 2.4\arcsec & 7 & 3.9\arcsec & 3.9\arcsec & 4 & -- & -- & -- \\
10-11hour & 2.1\arcsec & 2.1\arcsec & 16 & 2.8\arcsec & 2.8\arcsec & 150 & 3.8\arcsec & 3.8\arcsec & 83 & 4.3\arcsec & 4.0\arcsec & 6 \\
11-12hour & 17.0\arcsec & 4.0\arcsec & 0 & 2.1\arcsec & 2.1\arcsec & 2 & 3.7\arcsec & 3.7\arcsec & 1 & -- & -- & -- \\
12-13hour & 3.3\arcsec & 3.3\arcsec & 2 & 0.8\arcsec & -- & -- & 2.5\arcsec & 2.5\arcsec & 3 & -- & -- & -- \\
13-14hour & -- & -- & -- & 1.5\arcsec & -- & -- & 4.5\arcsec & 4.0\arcsec & 2 & -- & -- & -- \\
\end{tabular}
\label{tab: xmatch}
\end{table*}

\subsubsection{Results}
\label{sec: cross-match results}

We used the cross-match radii shown in Table\,\ref{tab: xmatch} to cross-match between each radio catalogue and each star catalogue. The results of the cross-matches are shown in the third sub-columns in Table\,\ref{tab: xmatch}. Note that there is some overlap between the results of the cross-matches. For example, a star might be found in the cross-match between RACS-high and CNS5 and that same star might have been found in the cross-match between RACS-low and the \textit{ROSAT} star catalogue. 

We performed final checks by checking the SIMBAD object type of each source.
HD 8357 (also known as AR Psc) and TWA 22 were both identified by Simbad as X-ray binaries. However, AR Psc is an RS CVn binary consisting of a K1 subgiant and a G7V dwarf \citep{1996AJ....112..269F,2022PASJ...74..477K} and TWA 22 is an M dwarf binary where both component are $\sim$ M5 stars \citep{2018A&A...618A..23R}. We have therefore kept both of these objects in the SRSC.

We removed Gaia DR3 247993683714250368 as it had an uncharacteristically high radio luminosity for a radio star: it is on the main sequence and is not a Wolf-Rayet star; however, its radio luminosity was calculated to be $\sim10^{19}\,\mathrm{erg\,s^{-1}\,Hz^{-1}}$. This is approximately two orders of magnitude higher than expected for this star, which we noticed as a clear outlier in luminosity in the CMD.
Only some of the Wolf-Rayet stars in the sample have similar luminosities. 
The radio spectrum from \citet{2021A&A...655A..17S} shows a power law spectrum with a spectral index of $\sim-0.7$. In this case, there is an optical star in front of a radio- and X-ray loud AGN.

We removed the Wolf-Rayet star CXOGC J174528.6$-$285605 after visually inspecting the ASKAP images, as the radio detection of this source appears to be slightly resolved.  
We also removed WR 76-6, WR 76-7 and WR 76-8 because these Wolf-Rayet systems were all matched to the same radio source as they are closely co-located on the sky and we could not distinguish which system was responsible for the radio emission.

There are two pairs of \textit{Gaia} sources: 

\begin{itemize}
    \item Gaia~DR3~4105057482987021184 and \\ Gaia~DR3~4105057482994688384;
    \item Gaia DR3 6130530322820978304 and \\ Gaia DR3 6130530322823774720
\end{itemize}

\noindent where both sources in the pair were identified as the same radio source due to the small separation between the \textit{Gaia} sources. Both pairs were identified as radio stars using the Stokes \textit{V} search method. After examining the \textit{Gaia} parameters for the pairs, both objects in each pair have near-identical proper motion and parallax values. This means that they are likely binary systems. As such we have kept these pairs of sources in the SRSC.

Performing these checks and removing the sources described above resulted in \srsccm\  unique stars.

\subsection{Other data and samples}
\label{sec: other data}

We have included in the SRSC the 37 radio stars in V-LoTSS found via circular
polarisation searches of LoTSS data \citep{vlotss} and 
the 22 stars found by cross-matching V-LoTSS and the GCNS from \citet{2023arXiv231207162Y}. 
We also included the 25 and 65 stars found by cross-matching LoTSS and VLASS, respectively, to the GCNS \citep{2023arXiv231207162Y}.

We have included stars found in searches for variable radio sources using ASKAP and MeerKAT. The ASKAP stars presented here were found by \citet{Yuanming_ShortTimescale} using the VAST short time scale imaging pipeline. They found eight stars with radio variability on minute timescales. Three MeerKAT-detected stars have been included in the catalogue: 

\begin{itemize}
    \item MKT~J170456.2$-$482100 \citep{Driessen_Flareyboi},
    \item EXO~040830$-$7134.7 \citep{Driessen_EXO0408}, and
    \item SCR~1746$-$3214 \citep{Andersson_Star}.
\end{itemize} 

\noindent All of these stars were found in untargeted searches for radio variability as part of the ThunderKAT project \citep{TKT_description}. We have included the eight stars found in proper motion searches by \citet{Driessen_ProperMotion}; two of the eight sources had not previously been identified as radio stars. 

\subsection{Samples that are not included}
\label{sec: data not included}

In this version of the SRSC we have included reliable identifications of radio stars from wide-field searches for radio stars using SKA precursors. We have not included candidate radio stars, such as the two tentative detections made by \citet{2018MNRAS.478.2835L} using the Murchison Wide Field Array \citep[MWA;][]{2012rsri.confE..36T}. 

We plan to perform a detailed literature search in the near future for detections from other searches and studies of radio stars, such as \citet{2020ApJ...904..138S}, \citet{2019ApJ...871..214V} and more. We plan to include these stars in a future version of the SRSC. Part of this literature search will include confirming the reliability of past detections, as some stars in e.g. the CRS were considered candidate detections but were still included as radio stars. We encourage submissions to the SRSC via email to the corresponding author.

We do not plan to include identification of stars in the mm-band, such as the detections using the South Pole Telescope \citep[SPT;][]{2021ApJ...916...98G,2016ApJ...830..143W} by \citet{2024arXiv240113525T}. This is because detections at these higher frequencies probe different physics to MHz/GHz radio detections. 

\section{Catalogue format}
\label{sec: catalogue format}

The SRSC consists of two tables, the \stab\ and the \rtab\ . The {\tt stars} table contains the identification information and basic properties of each unique star in the catalogue: there is one row per star. The \texttt{radio} table contains data for each radio measurement of each star in the catalogue: there may be more than one row per star. Both tables contain a unique identifier for each star of the format SRSC XXXXX to link the information in the tables together. The columns in the \texttt{star} table are described in Table\,\ref{tab: star table} and the columns in the \texttt{radio} table are described in Table\,\ref{tab: radio table}.

The astrometric information for the stars in the \stab\ is provided by the survey in the \texttt{survey} column. As described in Section\,\ref{sec: star cat prep}, the preferred survey for this information is \textit{Gaia} DR3; however, preference is given to the survey which includes proper motion information. We did not attempt to cross-identify sources between surveys based on source position because this is prone to misidentification. The different names for the stars are from SIMBAD, or were included if there was a cross-identification in the original survey information. For example, \textit{Gaia} source information does sometimes include the Tycho or 2MASS identification of a star. 

The \texttt{identification\_method} column in the \stab\ provides a binary flag to indicate which search methods were used to identify the star as a radio star.  This means that the user can filter the catalogue based on search method.
The \texttt{radio\_multiple} flag in the \stab\ indicates whether the radio-star cross-match has a one-to-one match or a one-to-many match. This is because the typical uncertainty on radio positions is larger than the separation between the components of some binary systems. ASKAP has typical astrometric accuracy of $1\arcsec$ - $2\arcsec$, LoTSS 0\farcs2, VLASS 0\farcs5, and MeerKAT 1\arcsec. This means that there may be one radio detection that is matched to both components of a binary system. Where there is a one-to-many match between one radio source and more than one star, we have set the \texttt{radio\_multiple} flag for the ``preferred'' star to 1 and we have set the flag for the ``duplicate'' star to 2. The ``preferred'' star is either the system identification, e.g. AT Mic instead of AT Mic A and AT Mic B, or the ``A'' component of the system. One-to-one matches have a \texttt{radio\_multiple} flag of 0. 

In the search described in Section \ref{sec: xmatch} we removed possible repeat detections to correctly calculate the cross-match reliability. However, since there are repeat ASKAP observations of many areas of sky, many of the stars are likely to have repeat detections.
Using the list of unique stars in the \stab, we searched all public ASKAP data up to 2024 March 26 to find repeat detections of these stars, using a match radius of 4\arcsec\. We have only included detections in the catalogue, not upper limits. All detections of \stab\ sources are in the \rtab. We show excerpts of the \stab\ (Figure\,\ref{tab: stars table demo}) and \rtab\ (Figure\,\ref{tab: radio table demo}) in \ref{app: table excerpts}.
 
The catalogue presented here is the first version of the SRSC. A static version of the catalogue is available in Vizier\footnote{The table will be uploaded to Vizier once the paper has been accepted to PASA. For now, the Vizier version of the catalogue can be downloaded from \href{https://radiostars.org/}{https://radiostars.org/}.}. It contains the stars presented in this paper. 
We will maintain and update a version of the catalogue on the website \href{https://radiostars.org/}{https://radiostars.org/}.

\begin{table*}
\caption{Description and units of the columns in the star table.}
\centering
\begin{tabular}{llp{13cm}}
Column name & Units & Description \\
\hline
Identifier &  & A unique identifier for each star in the SRSC. \\
SIMBAD &  & The main identifier from the CDS SIMBAD database \citep{simbad}. \\
Gaia &  & \textit{Gaia} unique source identifier \citep[unique within a particular Data Release,][]{GaiaMission,GaiaDR3,GaiaDR3_Validation}. For this release of the SRSC the \textit{Gaia} identifier is from DR3. . \\
Tycho &  & TYC1-3 (TYC number) Tycho identifier \citep{TYCHO_2,TYCHO_2_Validation}. \\
2MASS &  & 2MASS designation from the Two Micron All Sky Survey at IPAC \citep{2MASS}. \\
GCVS &  & Designation from the General Catalogue of Variable Stars (GCVS) \citep{GCVS}. \\
HIP &  & Hipparcos input catalogue running number \citep{hipparcos}. \\
UCAC4 &  & Recommended identifier from the Fourth U.S. Naval Observatory CCD Astrograph Catalog \citep{UCAC4}. \\
Survey &  & The survey used to provide the astrometric information shown here (i.e. the position, proper-motion and parallax information). \\
Survey\_id &  & Designation of the source in the survey used to provide the astrometric information for the source. \\
Epoch & jyear & The observation epoch of the source in the survey. The RA and Dec provided here is proper-motion correct/observed, all RA and Dec are in the J2000 reference frame. \\
RAdeg & degree & The J2000 RA in degrees proper-motion corrected to the date in the ``Epoch'' column. \\
e\_RAdeg & mas & The uncertainty on the Right Ascension. \\
DEdeg & degree & The J2000 Declination in degrees proper-motion corrected to the date in the ``Epoch'' column. \\
e\_DEdeg & mas & The uncertainty on the Declination. \\
plx & mas & The parallax provided by the survey. \\
e\_plx & mas & The uncertainty on the parallax. \\
pmRA & mas/yr & The proper-motion in Right Ascension ($\times\cos\left(\mathrm{Dec}\right)$). \\
e\_pmRA & mas/yr & The uncertainty on the proper-motion in Right Ascension ($\times\cos\left(\mathrm{Dec}\right)$). \\
pmDE & mas/yr & The proper-motion in Declination. \\
e\_pmDE & mas/yr & The uncertainty on the proper-motion in Declination. \\
Identification\_method &  & Indicates the search methods used to identify the radio emission as stellar emission from this star. The possible values are the sum of cross-match=2, variability=4, proper-motion=8, and Stokes V=16. For example, a source found in both a cross-match and Stokes V search would have a value of $2+16=18$. \\
Radio\_multiple &  & Flag indicating multiple optical matches to one radio source. 0 indicates that there is only one optical match to the radio component. Values of 1 and 2 indicate that there are multiple matches, where 1 is the ``preferred'' source and 2 is a ``duplicate''. This is for ease of use, such that selecting rows where this flag = 0 or 1 results in a set of one-to-one radio to optical matches. If one radio component has multiple optical matches this indicates that the optical matches are multiple components of the same stellar system. \\
\hline
\end{tabular}
\label{tab: star table}
\end{table*}

\begin{table*}
\caption{Description and units of the columns in the radio table.}
\centering
\begin{tabular}{llp{13cm}}
Column name & Units & Description \\
\hline
Identifier &  & A unique identifier for each star in the SRSC. \\
Telescope &  & The radio telescope used for the observation. \\
Survey &  & Name of the radio survey in which the star was identified. \\
Match\_separation & arcec & The separation between the radio and optical position, where the optical position has been proper motion corrected to the epoch of the radio observation. \\
Radio\_id &  & The identifier/name of the radio source in the radio survey. \\
Field &  & The designation/identifier of the radio field or observation in the radio survey. \\
Obs\_date & ISOT & ISOT start time of the radio observation. \\
Exposure & s & Integration time of the radio observation. \\
RAdeg & degree & The Right Ascension in degrees of the source in the radio observation. \\
DEdeg & degree & The Declination in degrees of the source in the radio observation. \\
e\_RAdeg & arcsec & The uncertainty on the Right Ascension. \\
e\_DEdeg & arcsec & The uncertainty on the Declination. \\
Freq & MHz & The central frequency of the radio observation. \\
SpeakI & mJy/beam & The peak flux density of the Stokes I detection of the source. \\
e\_SpeakI & mJy/beam & The uncertainty on the peak flux density of the Stokes I detection of the source. \\
StotI & mJy & The integrated flux density of the Stokes I detection of the source. \\
e\_StotI & mJy & The uncertainty on the integrated flux density of the Stokes I detection of the source. \\
bmax & arcsec & The major axis of the synthesised beam. \\
bmin & arcsec & The minor axis of the synthesised beam. \\
PA & degree & The position angle of the synthesised beam. \\
e\_bmax & arcsec & Uncertainty on the major axis of the synthesised beam. \\
e\_bmin & arcsec & Uncertainty on the major axis of the synthesised beam. \\
SpeakV & mJy/beam & The peak flux density of the Stokes V detection of the source. \\
e\_SpeakV & mJy/beam & The uncertainty on the peak flux density of the Stokes V detection of the source. \\
StotV & mJy & The integrated flux density of the Stokes V detection of the source. \\
e\_StotV & mJy & The uncertainty on the integrated flux density of the Stokes V detection of the source. \\
localrmsV & mJy/beam & The local Stokes I rms. \\
localrmsI & mJy/beam & The local Stokes V rms. \\
Ref &  & The reference (bibcode) for the radio star identification. \\
\hline
\end{tabular}
\label{tab: radio table}
\end{table*}

\section{Properties of radio stars in the SRSC}
\label{sec: catalogue properties}

\begin{figure*}
\caption{Sky map showing the positions of the SRSC stars in Galactic coordinates. The stars are coloured by their maximum radio luminosity. The black dashed line indicates Declination$=0^{\circ}$ and the black dotted line indicates Declination$=+40^{\circ}$.
}
\includegraphics[width=\columnwidth]{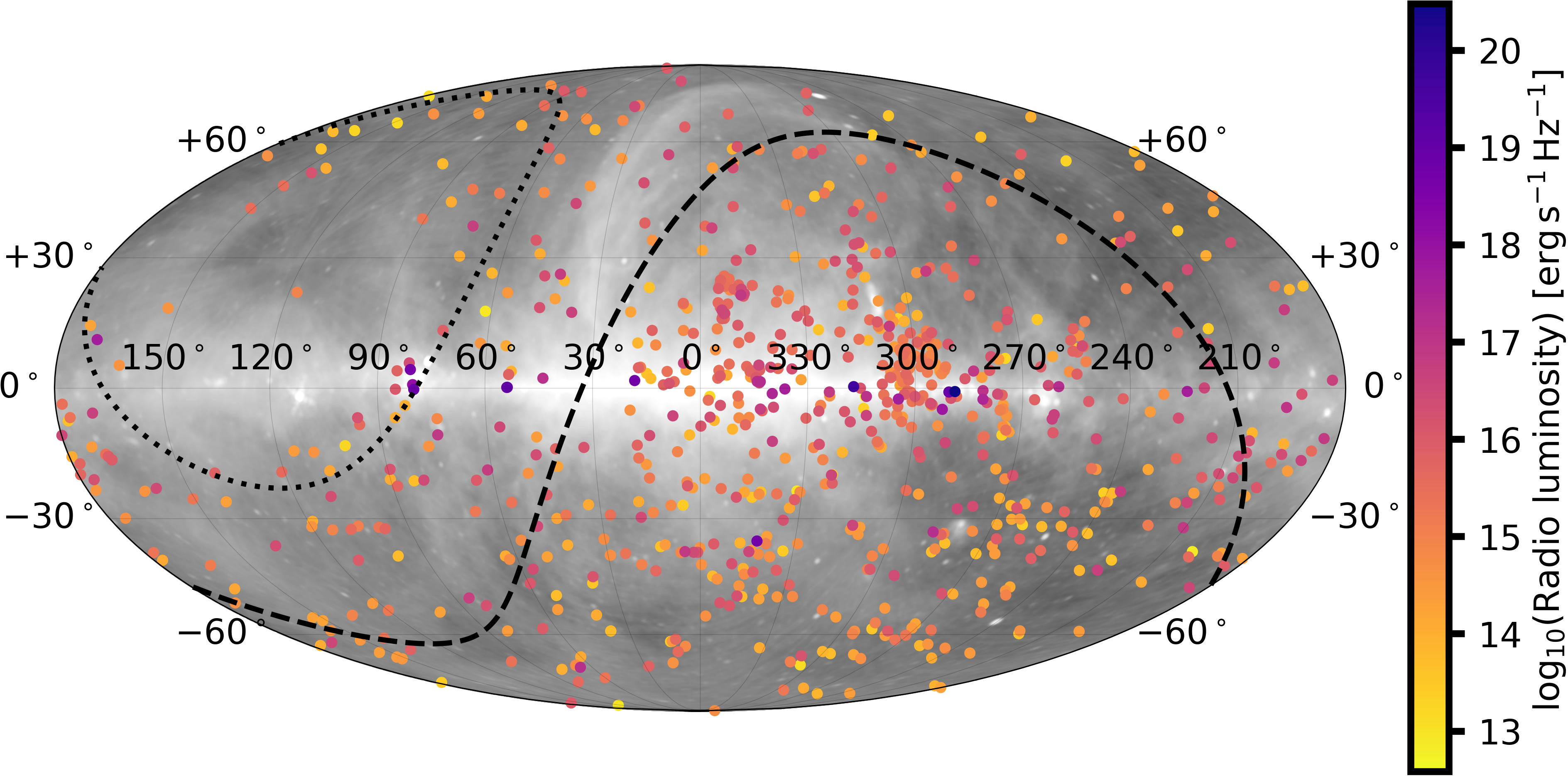} % switch to pdf version before submitting, it just takes ages to load
\label{fig: SkyMap}
\end{figure*}

The current version of the SRSC comprises \srsctot\ unique stars 
with \srscdet\ total radio detections (\srscsingles\ stars have a 
single radio detection while \srscmultis\ have more than one) at frequencies
from $144$\,MHz to $3$\,GHz. The sky distribution of 
the stars is shown in Figure\,\ref{fig: SkyMap}. 
The stars are distributed across the whole sky; 
with $\sim74$ per cent at Dec$<0^{\circ}$.
This is because nearly 
75 per cent of ASKAP observations are of this region and $\sim95$ per cent of the stars in the SRSC have been detected at least once using ASKAP. 
The integrated flux densities over all radio detections 
range from 0.02\,mJy (EXO\,040830$-$7134.7) to 199.4\,mJy 
(Apep). 

\begin{figure}
\caption{Number of sources detected using the cross-match (CM), circular polarisation (CP) and proper motion and variability (PM \& V) methods.
}
\includegraphics[width=\columnwidth]{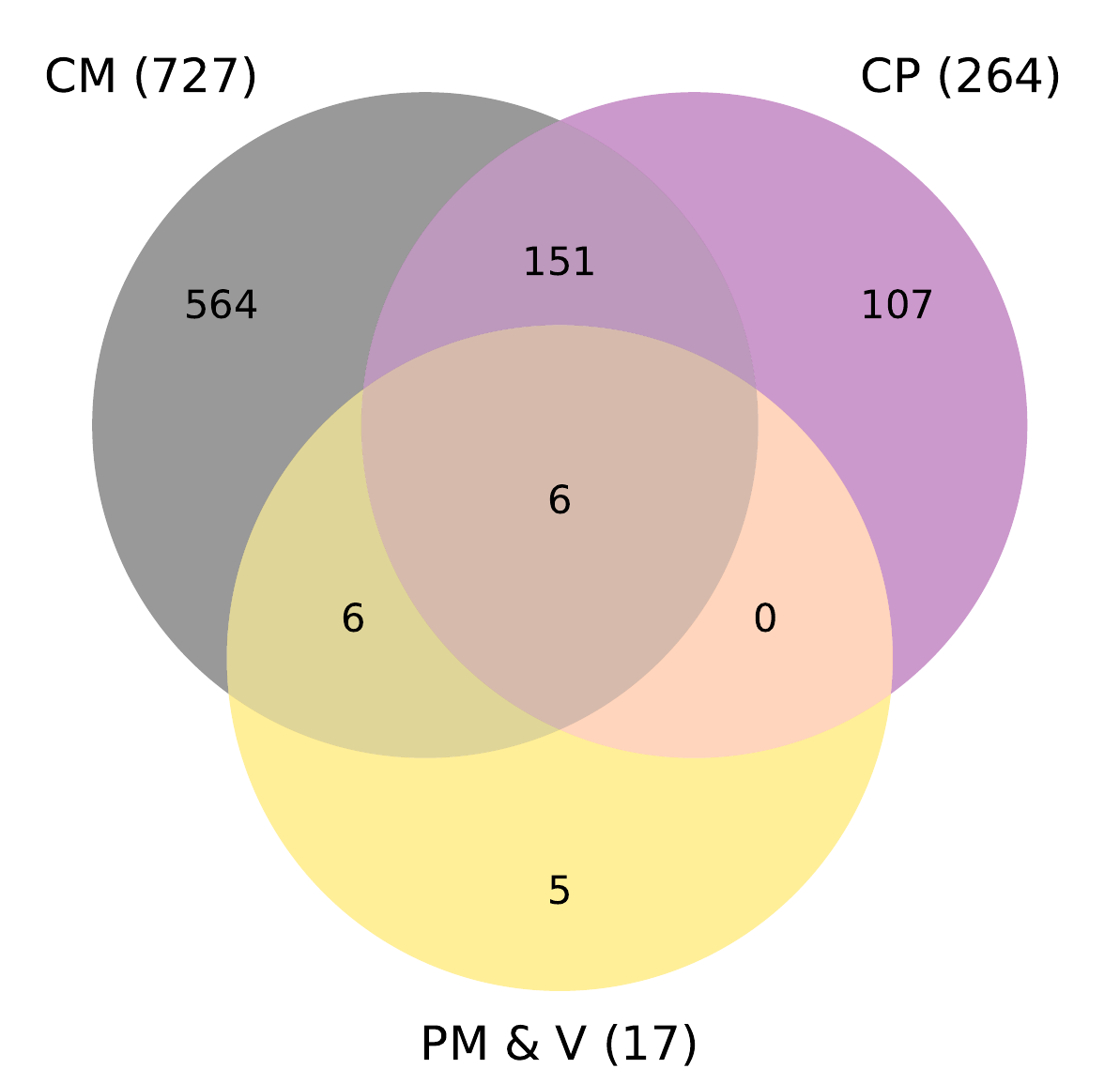}
\label{fig: Count venn}
\end{figure}

The number of stars detected using the different methods is shown in Figure\,\ref{fig: Count venn}, with the proper motion detected (8 total) and variability detected (9 total) stars combined into one set due to their comparatively low numbers. The effectiveness of each method is determined by a combination of factors.
For example: circular polarisation searches typically find coherent radio emission and flares while the cross-match searches included longer integration images so are more likely to find fainter, quiescent radio emission.

A distance measurement is needed to calculate the luminosities 
of the stars in the SRCS. We used the \texttt{rgeo} distance measurements
from \citet{2021AJ....161..147B} to calculate the luminosities for the SRSC stars.
\srscdist\ of the \srsctot\ stars have an \texttt{rgeo} distance measurement.
The nearest star is Proxima Centauri (at 1.3\,pc) and the most distant
star is WRAY 15-682 (also known as TYC 8958-1166-1, at 7315\,pc), both of which  have been detected in the radio before \citep{1978ApJ...225L..35H,2002MNRAS.330...63D}.

\subsection{Colour-magnitude diagram}
\label{sec: CMD}

\begin{figure*}
\caption{\textit{Gaia} DR3 CMD showing the radio stars in the SRSC in colour. The colour scale shows the radio luminosity based on the maximum flux density of each star in the SRSC and the \textit{Gaia} \texttt{rgeo} distance. The grey background points show the \textit{Gaia} DR2 CMD for reference \citep{2019ApJ...872L...9P}. 
}
\includegraphics[width=\columnwidth]{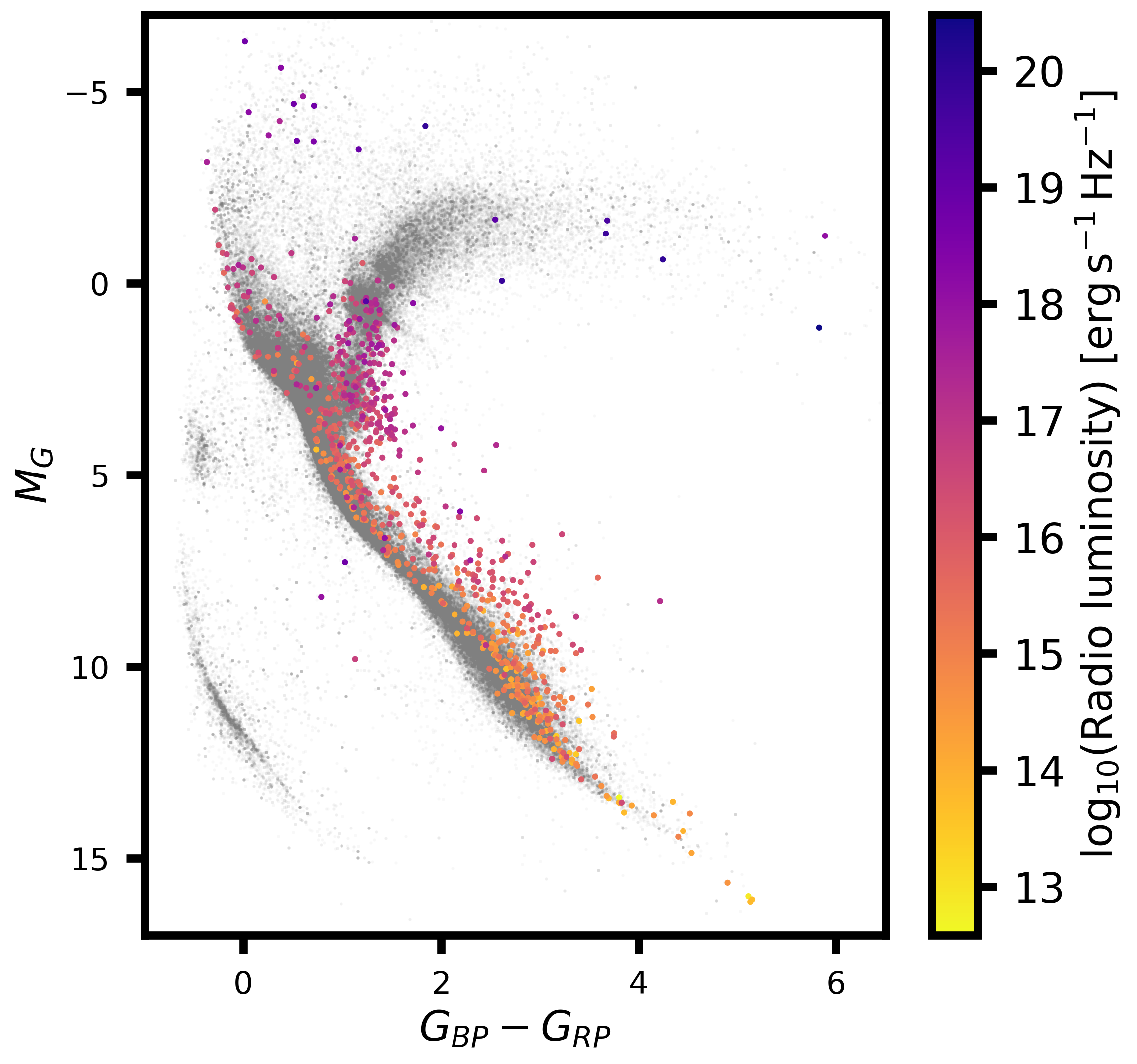} % switch to pdf version before submitting, it just takes ages to load
\label{fig: CMD luminosity}
\end{figure*}

Colour-magnitude diagrams (CMDs) are a useful tool 
for giving an overview of the types of stars in a 
sample. We used the set of stars that have \textit{Gaia} 
DR3 magnitudes and \texttt{rgeo} distances to create a CMD of the 
stars in the SRSC, shown in Figure\,\ref{fig: CMD luminosity}. $M_{G}$ is the absolute G-band ($\sim$330-1050 nm) magnitude and $G_{BP}$ and $G_{RP}$ are the \textit{Gaia} blue (330-680 nm) and red (640-1050 nm) apparent magnitudes. 
Figure\,\ref{fig: CMD luminosity} shows that there are radio detected stars across the CMD. We can see that stars that have a brighter $M_{G}$ are also more radio luminous. There is a large cluster of stars on the sub-giant part of the red giant branch and some stars are in the red clump. These sub-giant stars are more radio luminous than the main sequence stars. 
We can also see that many of the stars are above the main sequence. Many of these are known to be pre-main-sequence stars, such as T~Tauri stars and Orion variables. These stars tend to be more radio luminous than the stars on the main sequence at a similar $M_{G}$.
The cool dwarfs, the stars on the main sequence with the faintest $M_{G}$, also have the lowest radio luminosities. This means that we are detecting a closer-by sample than the sample of sub-giant and giant stars that are brighter in the radio.

There are ten objects in the SRSC that do not have a \textit{Gaia} identifier from any data release. Sources are not in \textit{Gaia} when they are either too faint (e.g. ultracool dwarfs),  too bright, or in a crowded region. Three of the ten objects without \textit{Gaia} identifiers are ultracool dwarfs: 2MASS J09481615$+$5114518, 2MASS J10534129$+$5253040 and [VCS2020] BDR J1750$+$3809.
The other seven objects include the bright star (magnitude $\sim1.5$) Castor AB and the Wolf-Rayet stars Regor, [KSF2015] 1256$-$1483A, and Apep that are too bright to be in \textit{Gaia}. The remaining objects are UPM J1709$-$5957,\\ 1RXS J200031.8$+$592127 and V503 Hya.
It may be that they are binary systems so the individual stars have \textit{Gaia} identifiers but not the system. Or there may be two \textit{Gaia} sources close to the star and so they have not been linked to either of them in SIMBAD or other catalogues.

\subsubsection{M dwarfs}
\label{sec: mdwarfs}

\begin{figure*}
\caption{Light curves of the M dwarfs in the SRSC that have ten or more radio detections. MJD 58500 is 2019 January 17.
}
\includegraphics[width=0.96\columnwidth]{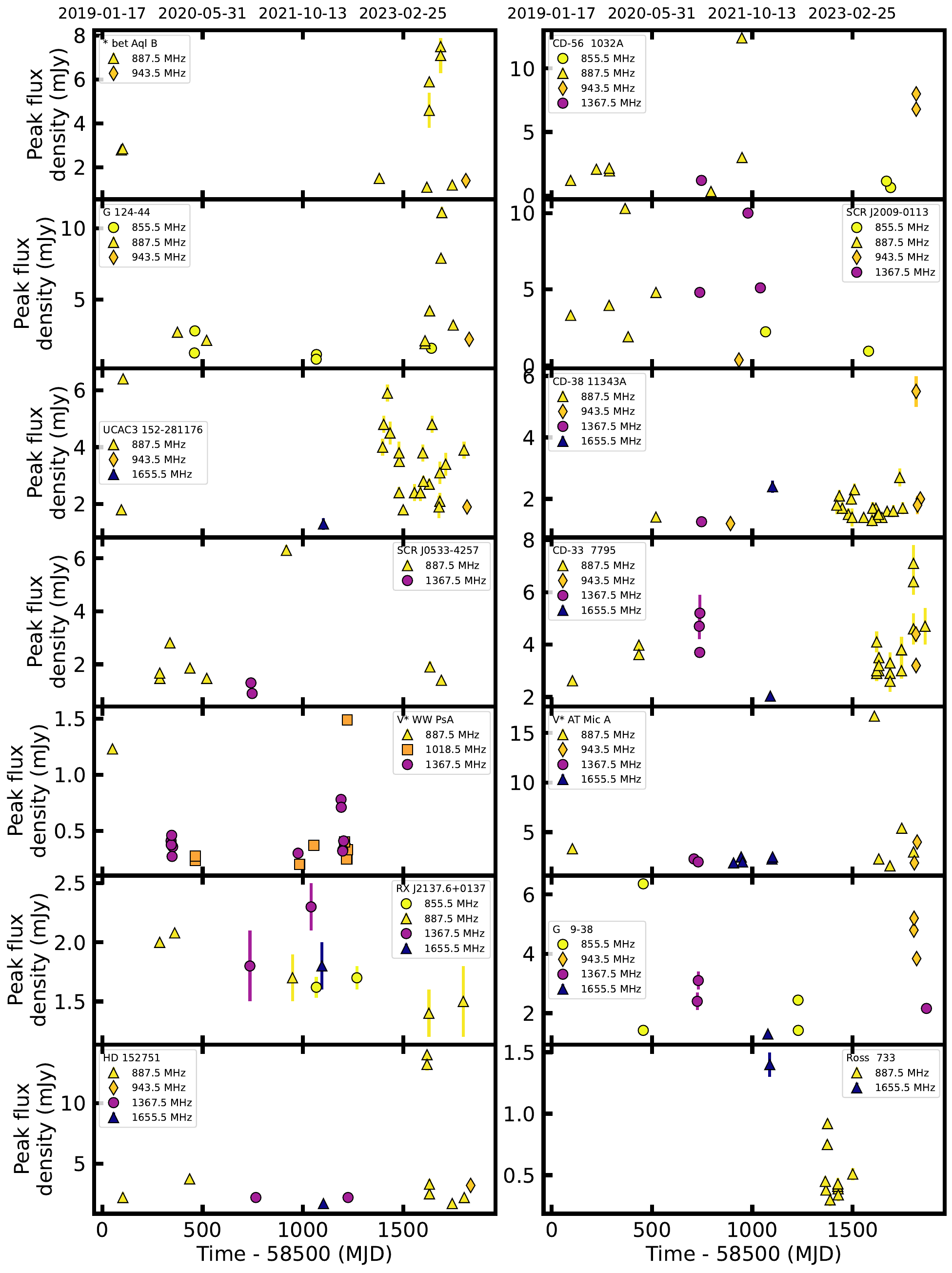} 
\label{fig: Mdwarf light curves}
\end{figure*}

M dwarfs are one of the most common types of detectable radio star. There are 208 stars in the SRSC that are classified in SIMBAD as either a single star of ``M'' spectral type, or a binary with at least one ``M'' spectral type component. Just under half of these stars, 106, are detected in the radio more than once. Since there are so many M dwarfs with multiple detections we are unable to show all the light curves here. We show the light curves of the fourteen M dwarfs with ten or more radio detections in Figure\,\ref{fig: Mdwarf light curves}. We can see that these stars are highly variable in the radio, with some higher flux density detections that may indicate bursts or flares.

\subsubsection{Stars below the main sequence}
\label{sec: wds}

There are three stars in the SRSC that fall below the main-sequence in the CMD in Figure\,\ref{fig: CMD luminosity}: the cataclysmic binary 2MASS J19551247$+$0045365 \citep[also known as IGR J19552$+$0044;][]{2010A&A...519A..96M,2013MNRAS.435.2822B}, and \textit{Gaia} sources Gaia DR3 1846427022035230592 and~ 961025698716736768. Stars in this area of the CMD are often binary systems with a white dwarf component or B-type subdwarfs. 

2MASS J19551247$+$0045365 is an asynchronous short-period polar type CV \citep{2017A&A...608A..36T}. The white dwarf component has a magnetic field below 20\,MG and a spin period of $81.29\pm0.01$\,m, while the binary system has an orbital period of $83.599\pm0.002$\,m.
2MASS J19551247$+$0045365 is a known radio source \citep{2020AdSpR..66.1226B}. \citet{2020AdSpR..66.1226B} observed it using the VLA X-band (8-12\,GHz) and detected a flux density of $0.073\pm0.007$\,mJy. We detect the source twice with flux densities of $2.2\pm0.03$ and $1.6\pm0.2$\,mJy at 887.5 and 943.5\,MHz respectively.

The other two sources are not well studied and have not previously been identified as radio stars. Further investigation is required to identified their stellar type and determine if they are white dwarf binary systems.
All three of these sources were found in the circular polarisation search described in Section\,\ref{sec: ASKAP circ pol search}. Gaia DR3 1846427022035230592 and Gaia DR3 961025698716736768 have each only been detected once in the radio, in RACS-high, with flux densities $1.7\pm0.2$\,mJy and $1.1\pm0.2$\,mJy respectively. 

\subsubsection{Luminous radio stars and Wolf-Rayet systems}
\label{sec: WRs}

\begin{figure*}
    \caption{\label{fig: WR light curves} Radio light curves of the luminous radio stars and Wolf-Rayet stars in the SRSC that have been detected three or more times.  MJD 58500 is 2019 January 17.}
\includegraphics[width=\columnwidth]{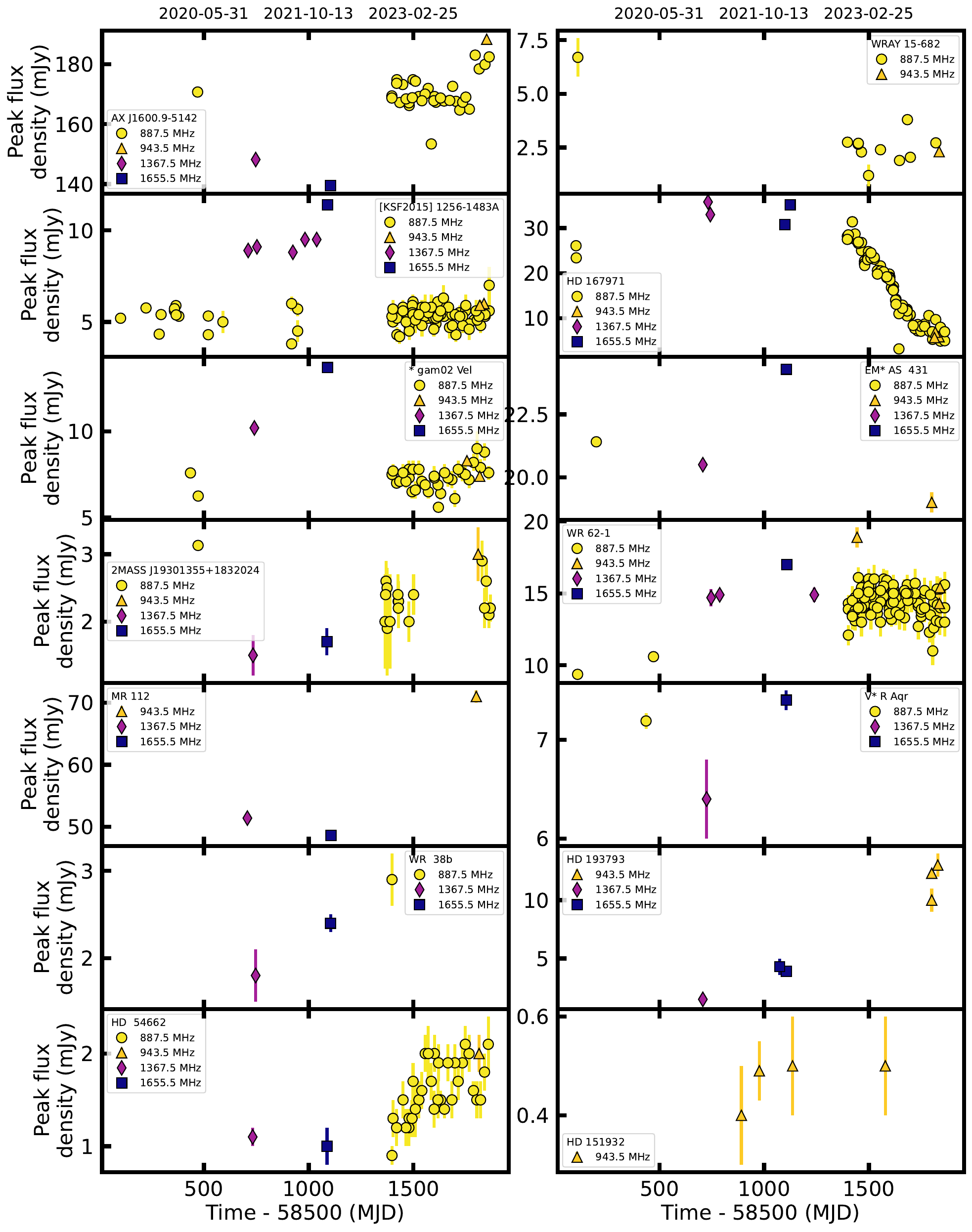}
\end{figure*}

The most radio luminous stars are in the top left of the CMD (brighter $M_{G}$) as well as the stars on the giant branch. These stars include Wolf-Rayet stars, symbiotic binaries, and blue supergiants. We selected these stars by choosing stars where the SIMBAD main classification is ``WolfRayet*'', or where $G_{BP} - G_{RP} > 2.5$ and $M_{G} < 2$, or where $M_{G} > 2$. There are a total of 24 stars that meet one or more of these criteria and fourteen of those stars have been detected three times or more in the radio. The light curves of the fourteen luminous radio stars that have been detected three or more times are shown in Figure\,\ref{fig: WR light curves}. 
We can see that these stars are highly variable. Many of these stars are within the VAST footprint, and as such have been detected many times. 
HD~167971 (also known as MY Ser) has a particularly interesting light curve. MY Ser is a hierarchical triple system at a distance of 1.8\,kpc. It consists of a 3.3\,day binary (O7.5III$+$9.5III) with a third component in a 21.4\,year orbit. It is a well-known radio source. It has been observed to have long-term radio variability on time scales of years \citep[e.g.][]{2007A&A...464..701B}. We can see in its light curve that the flux density of MY Ser has been steadily declining over approximately a year. 
The ten stars that are detected only once or twice are 
$\lambda$ Lep,
$\theta$ Mus,
HD 148937,
HD  86161,
HD  96548,
Schulte  9,
HD 152408,
IPHAS J193038.84+183909.8,
HD 156385
and
HD 152270.

\subsection{\gudel--Benz relation}
\label{sec: gb}

The \gudel--Benz relation is the correlation between the radio (5-9\,GHz) and X-ray (soft) luminosity of chromospherically active stars \citep{1993ApJ...405L..63G}.
Past investigations of this relation have typically demonstrated it using a few tens of stars \citep[e.g.][]{2021NatAs...5.1233C,2022ApJ...926L..30V,2023arXiv231207162Y}.
The SRSC more than doubles the known number of radio stars, and it is important to complement it with a similarly sensitive large scale X-ray survey. We have used eROSITA soft X-ray data to find the X-ray detections of these objects.

eROSITA \citep{2021A&A...647A...1P} is an X-ray instrument onboard the SRG spacecraft, that performed a deep all-sky survey in the 0.2\,--\,10.0~keV energy range.
The eROSITA all-sky survey (eRASS) data is accumulated in successive all-sky surveys with a duration of about 0.5~yr each. The DR1 data release of the eRASS1, that contains the first half-year of data from the western Galactic hemisphere and a description of the survey, instrument performance as well as data products and source catalogues are presented in \cite{eROSITA_DR1}.
For our analysis we use the X-ray catalogue from the more sensitive, accumulated data from the first four all-sky surveys (eRASS:4, 2~yr data, version c020/221031) and perform a positional cross-match with the SRSC sample, using a matching radius of 10\arcsec. The matching radius is adapted to the average positional uncertainty of eROSITA, while likewise keeping the number of false positives at an acceptable level, in this case below 1\%. Note that also this eRASS catalogue covers only half of the sky, roughly $l\geq180\,$deg. The adopted X-ray fluxes are survey averages and refer to the standard eROSITA energy band of 0.2\,--\,2.3~keV.

\begin{figure*}
\caption{\gudel--Benz plot showing the radio stars in the SRSC. The X-ray luminosity is from eROSITA. 
We show the stars that were found using the circular polarisation method (CP), and the stars that were found using other methods (other).
Each star is only present once. The grey-dashed line shows the \gudel--Benz relation from \citet{2014ApJ...785....9W}.
}
\includegraphics[width=\columnwidth]{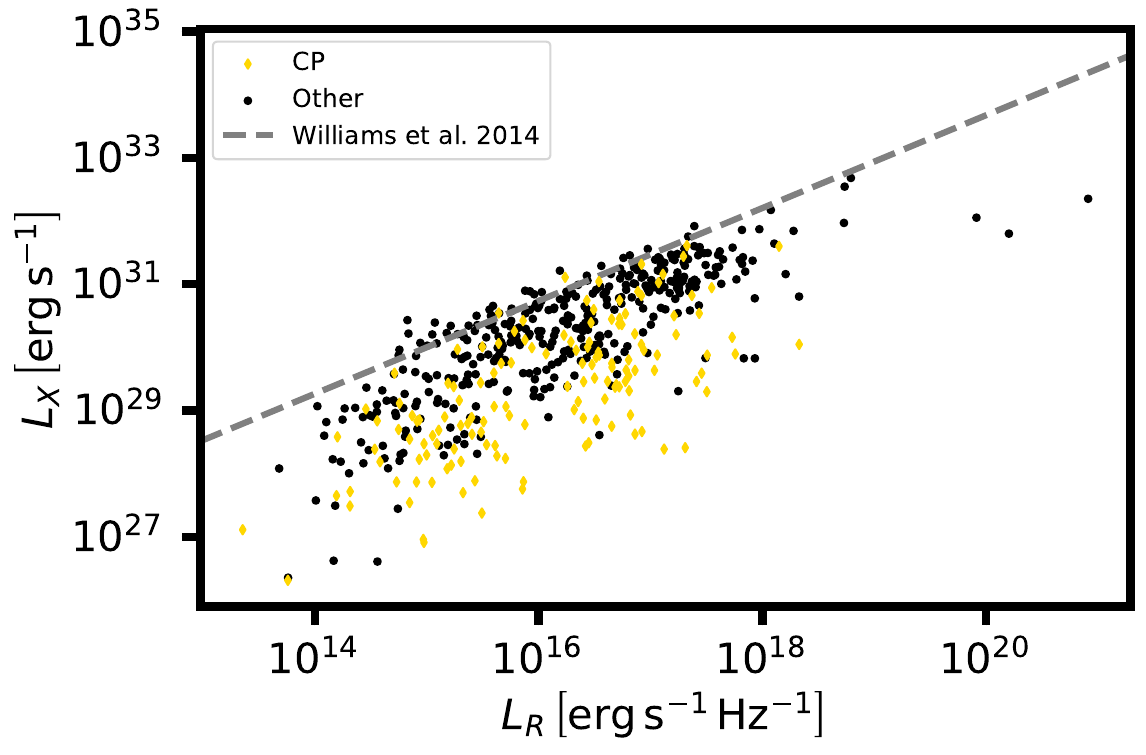}
\label{fig: GB methods}
\end{figure*}

We show the SRSC radio luminosities and eROSITA X-ray luminosities for the SRSC stars in Figure\,\ref{fig: GB methods}. The western Galactic hemisphere is largely in the southern sky, the same area that is well covered by ASKAP. There are \srscdist\ stars in the SRSC with \textit{Gaia} \texttt{rgeo} distance measurements, \srscero\ of those have eROSITA X-ray detections. We can see that most stars in the SRSC are over-luminous in the radio (or under-luminous in the X-ray) compared to the expected correlation. This has been noted before using both LOFAR ($\sim$200 MHz) and VLASS (3 GHz). \citet{2023arXiv231207162Y} showed that the radio stars they found by cross-matching LOFAR and VLASS to the GCNS are consistently over-luminous in the radio (or under-luminous in the X-ray) compared to the previous \gudel--Benz relation. \citet{2021NatAs...5.1233C} similarly found that the M dwarfs they observed with LOFAR could be orders of magnitude over-luminous in the radio when compared to the canonical \gudel--Benz relation. This is because radio emission from stars at lower frequencies is more likely to be coherent radio emission \citep{1990SoPh..130..265B}, which is more luminous and is not expected to contribute to heating the corona (which contributes to the X-ray emission). Figure\,\ref{fig: GB methods} is further evidence of this.
However, \citet{2022ApJ...926L..30V} and \citet{2023arXiv231207162Y} found a population of chromospherically active, coherent radio emitting stars that follow the relation. It is unclear why these stars adhere to the \gudel--Benz relation at lower frequencies.

The original \gudel--Benz relation was demonstrated using radio detections of stars at 5 and 9 GHz and X-ray detections with \textit{ROSAT}. The eROSITA energy band is similar to the \textit{ROSAT} energy band, but the SRSC currently only includes radio detections below 3\,GHz, with most detections from ASKAP between 800 and 1700\,MHz. We plan to explore the dependence of the \gudel--Benz relation on frequency and whether there are different correlations for different stellar types at different frequencies, or if there is a de-coupling of the relation at lower frequencies. Simultaneous or quasi-simultaneous observations at a wide range of radio frequencies will show if there is a spectral change between 3 and 5\,GHz that would indicate a change in the radio mechanism from coherent radio emission to incoherent gyrosynchrotron. 
The transition frequency between two emission regimes – e.g. non-thermal to thermal, or incoherent to coherent -- encodes information about the conditions in the stellar coronae, such as maximum magnetic field strength or particle density.

% The Australia Telescope Compact Array \citep[ATCA,][]{1992JEEEA..12..103F} will be upgraded to the Broadband Integrated GPU Correlator for ATCA (BIGCAT) system in the near future. BIGCAT will enable observations with a contiguous 7.68\,GHz bandwidth, which  will help better understand the spectra of radio stars and improve our understanding of the radio and X-ray emission mechanisms and any links between them. We also plan to expand the SRSC by cross-matching to X-ray emitting stars identified using eROSITA \citep{2024arXiv240117282F}. 

\subsection{Comparison with other radio star catalogues}
\label{sec: comparison to catalogues}

The Catalogue of Radio Stars \citep[CRS;][]{Wendker_1978,Wendker_1987,Wendker_1995} was constructed by performing literature reviews and included both radio detections and non-detections of stars. In some cases, radio stars were included when a radio source was detected in a dense cluster of stars. In most other cases it included radio stars that were found in targeted searches for stellar radio emission. In this work we have chosen to include radio detections of stars found in large scale searches for radio stars using SKA precursor instruments and a variety of different methods. This means that the current version of the SRSC will be missing some of the stars in the CRS, including some well known radio stars like BY Draconis and P Cygni. 

We have not incorporated the CRS directly into the SRSC because of the inconsistent methods of detection and the challenge of determining the reliability of some of the radio star associations. We plan to evaluate the reliability of detections in the CRS and incorporate them, as part of future work. 

\begin{figure}
\caption{Histogram of the detection frequency of radio stars in the CRS and SRSC. We have only showed the CRS stars detected below 10\,GHz here (766 stars). Each star is only counted once. For the CRS stars we include the lowest frequency detection of each star. For the SRSC stars we include the highest frequency detection of each star. We chose to show the frequencies detected this way to demonstrate that even showing the closest set of frequencies, the SRSC contains many more lower frequency detections than the CRS.
}
\includegraphics[width=\columnwidth]{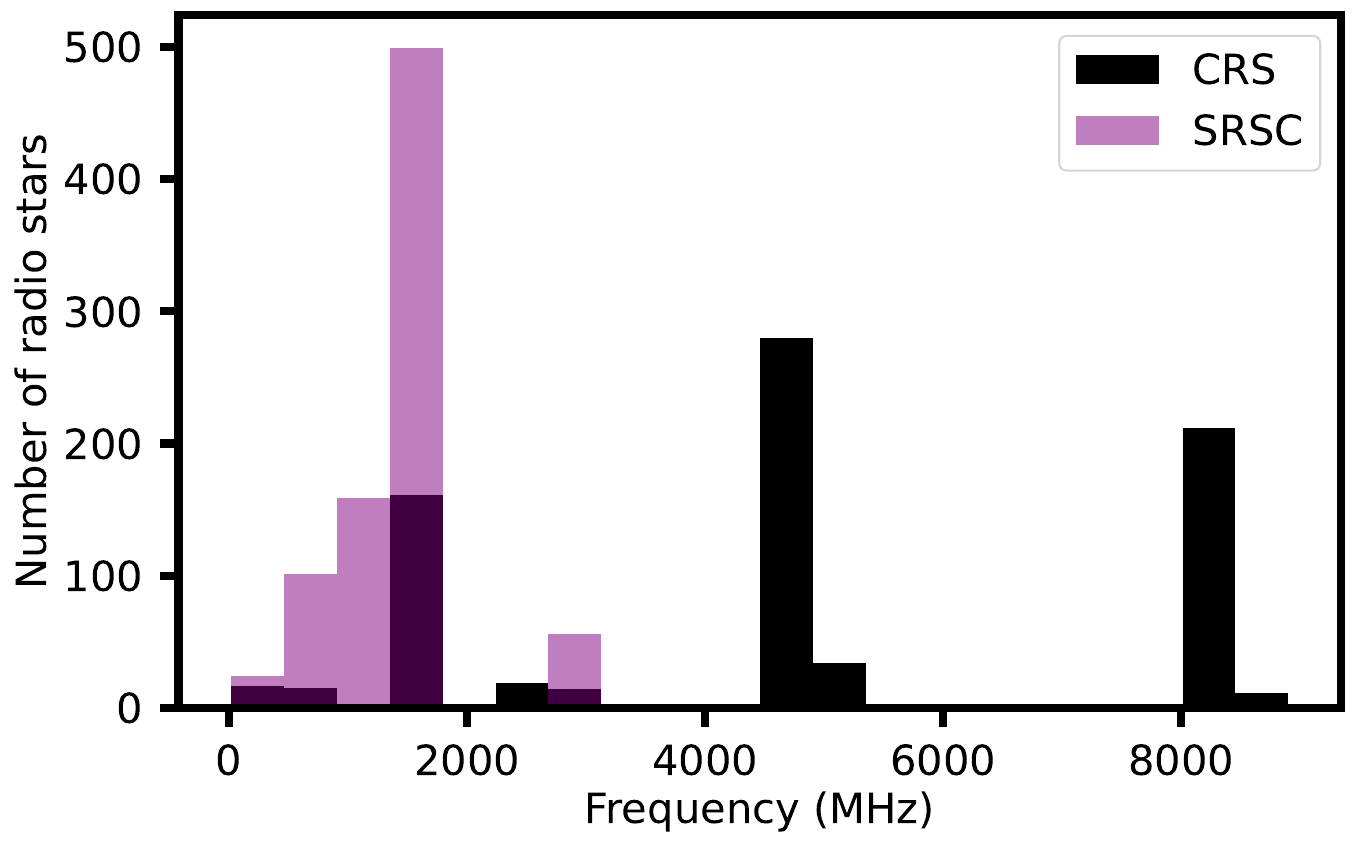}
\label{fig: crs srsc hist}
\end{figure}

We can, however, compare the number of stars in the CRS to the number of stars in the SRSC.
The CRS \citep{Wendker_1978,Wendker_1987,Wendker_1995} 
contains 1042 stars detected at frequencies below 
420.0\,GHz, 
228 of which were detected at frequencies below 3\,GHz. 
The SRSC only contains stars detected at frequencies below 
3\,GHz. We show a histogram of the number of stars detected 
at various frequencies in the CRS and SRSC in 
Figure\,\ref{fig: crs srsc hist}. 
127 SRSC stars are in the CRS, 70 of which were detected below 3\,GHz in the CRS.

Other searches for radio stars include \citet{2001ApJS..135..227B} and \citet{2009ApJ...701..535K}. \citet{2001ApJS..135..227B} used a 1\,arcsec\ radius to cross-match between FIRST radio sources and optical sources from the Cambridge Automated Plate Measuring Machine (APM) scans of the POSS I plates. They classified sources as ``stellar'' if the optical source was unresolved and estimated that 98 per cent of the associations they found were physically associated. They found 107 stellar and unknown sources. \citet{2009ApJ...701..535K} used a 1\,arcsec\ radius to cross-match between FIRST radio sources and Sloan Digital Sky Survey \citep[SDSS;][]{2009ApJS..182..543A} sources. They found 112 matches, but concluded that most of their matches were chance coincidence after filtering and visual inspection. As both of these searches used FIRST data, they only include sources above $-10^{\circ}$ declination. Although 286 sources in the SRSC are above $-10^{\circ}$ declination, we do not find any matches between the SRSC and the sources found by \citet{2001ApJS..135..227B} and \citet{2009ApJ...701..535K}.
The majority of stars in the SRSC were found via cross-matching (see Figure\,\ref{fig: Count venn}), and none of the stars from \citet{2001ApJS..135..227B} or \citet{2009ApJ...701..535K} are in the optical catalogues (see Table\,\ref{tab: Optical catalogue lengths}) we cross-matched to. Only 44 of the stars in the SRSC found using the circular polarisation search method (and not via cross-matching) have a Declination $>-10^{\circ}$. These low numbers mean that it is not surprising that we did not find any sources that were found by \citet{2001ApJS..135..227B} and \citet{2009ApJ...701..535K}. Future searches for radio stars using the VLA and LOFAR, for example, may reveal whether any of the \citet{2001ApJS..135..227B} and \citet{2009ApJ...701..535K} radio star identifications are reliable.

\subsection{Future work}
\label{sec: future work}

We aim to continue to update the SRSC as more radio stars are identified. To that end, we plan to search for radio detections of eROSITA-identified X-ray emitting stars \citep{2024arXiv240117282F}, VAST-identified variable stars and more. 

We also plan to perform a detailed literature review to include past identifications of radio stars, including only those that are confirmed radio stars and excluding candidate radio stars or sources that are now known to not be radio stars. We will include stars that are detected by both SKA precursors and other radio telescopes. We will also include confirmed radio stars found by other teams. Please email the corresponding author if you wish to submit radio stars to the SRSC.

The SRSC, particularly as it is expanded and updated in the lead up to the SKA, will be an important resource for identifying radio stars for SKA investigation as well as for predicting the expected population of radio stars, and identifying properties of stellar radio emission so new radio stars can be identified. The SRSC will also be useful for cosmological applications which would benefit from the removal of Galactic foreground sources, for example cosmic microwave background dipole studies \citep[e.g.][]{2021A&A...653A...9S,2023A&A...675A..72W}. 

\section{Summary}
\label{sec: summary}

We have presented the SRSC, a catalogue of \srsctot\ radio stars identified using circular polarisation searches, cross-matching (with 98\% reliability), variability searches and proper motion searches. This version of the SRSC includes stars found using SKA precursor instruments, and we plan to update the catalogue to include past and future identifications of radio stars. We have demonstrated the science potential of the SRSC using the CMD, radio light curves of a small sample of radio stars and the \gudel--Benz relation.

\section*{Acknowledgements}
The authors would like to thank Tin Lai for his work on the radio stars catalogue website.
LND would like to thank Courtney Crawford, May Gade Pedersen and Tim Bedding for helpful discussions about Hertzsprung-Russell and colour-magnitude diagrams.

DK is supported by NSF grant AST-1816492.

BJSP acknowledges the traditional owners of the land on which the University of Queensland is situated, upon whose unceded, sovereign, ancestral lands he works, and pays respect to their Ancestors and descendants, who continue cultural and spiritual connections to Country. 

LND would like to acknowledge the Gadigal People of the Eora nation, the traditional owners of the land on which most of this work was completed.

KR thanks the LSST-DA Data Science Fellowship Program, which is funded by LSST-DA, the Brinson Foundation, and the Moore Foundation; their participation in the program has benefited this work.

This scientific work uses data obtained from Inyarrimanha Ilgari Bundara / the CSIRO's Murchison Radio-astronomy Observatory. We acknowledge the Wajarri Yamaji People as the Traditional Owners and native title holders of the Observatory site. CSIRO’s ASKAP radio telescope is part of the \href{https://ror.org/05qajvd42}{Australia Telescope National Facility}. Operation of ASKAP is funded by the Australian Government with support from the National Collaborative Research Infrastructure Strategy. ASKAP uses the resources of the Pawsey Supercomputing Research Centre. Establishment of ASKAP, Inyarrimanha Ilgari Bundara, the CSIRO Murchison Radio-astronomy Observatory and the Pawsey Supercomputing Research Centre are initiatives of the Australian Government, with support from the Government of Western Australia and the Science and Industry Endowment Fund.

This paper includes archived data obtained through the CSIRO ASKAP Science Data Archive, 
CASDA\footnote{\href{http://data.csiro.au/}{http://data.csiro.au/}}.

This work has made use of data from the European Space Agency (ESA) mission
\href{https://www.cosmos.esa.int/gaia}{\it Gaia}, processed by the {\it Gaia}
Data Processing and Analysis Consortium (\href{https://www.cosmos.esa.int/web/gaia/dpac/consortium}{DPAC}). Funding for the DPAC
has been provided by national institutions, in particular the institutions
participating in the {\it Gaia} Multilateral Agreement.

This work uses data from eROSITA, the soft X-ray instrument aboard SRG, a joint Russian-German science mission supported by the Russian Space Agency (Roskosmos), in the interests of the Russian Academy of Sciences represented by its Space Research Institute (IKI), and the Deutsches Zentrum f{\"u}r Luft- und Raumfahrt (DLR). The SRG spacecraft was built by Lavochkin Association (NPOL) and its subcontractors, and is operated by NPOL with support from the Max Planck Institute for Extraterrestrial Physics (MPE).
The development and construction of the eROSITA X-ray instrument was led by MPE, with contributions from the Dr. Karl Remeis Observatory Bamberg \& ECAP (FAU Erlangen-N{\"u}rnberg), the University of Hamburg Observatory, the Leibniz Institute for Astrophysics Potsdam (AIP), and the Institute for Astronomy and Astrophysics of Universit{\"a}t T{\"u}bingen, with the support of DLR and the Max Planck Society. The Argelander Institute for Astronomy of the University of Bonn and the Ludwig Maximilians Universit{\"a}t M{\"unchen} also participated in the science preparation for eROSITA. The eROSITA data shown here were processed using the eSASS software system developed by the German eROSITA consortium.

This research made use of \href{http://www.astropy.org}{Astropy}, a community-developed core Python package for astronomy \citep{2013A&A...558A..33A,2018AJ....156..123A}; andAPLpy, an open-source plotting package for Python \citep{2012ascl.soft08017R}.

This research has made use of the \href{https://doi.org/10.26093/cds/vizier}{VizieR} catalogue access tool \citep{2000A&AS..143...23O} and the SIMBAD Database \citep{2000A&AS..143....9W}, both operated at the CDS, Strasbourg, France.

This publication made use of data products from the Two Micron All Sky Survey, which is a joint project of the University of Massachusetts and the Infrared Processing and Analysis Center/California Institute of Technology, funded by the National Aeronautics and Space Administration and the National Science Foundation.

%%%%%%%%%%%%%%%%%%%%%%%%%%%%%%%%%%%%%%%%%%%%%%%%%%

\section*{Data Availability}

The Sydney Radio Star Catalogue is available on Vizier or at \href{https://radiostars.org/}{https://radiostars.org/}.
ASKAP data are available in the CSIRO ASKAP Science Data Archive (CASDA)\footnote{\href{https://research.csiro.au/casda/}{https://research.csiro.au/casda/}}. The \textit{eROSITA}-DE Data Release 1 is available in Vizier \citep{eROSITA_DR1}\footnote{\href{https://cdsarc.cds.unistra.fr/viz-bin/cat/J/A+A/682/A34}{https://cdsarc.cds.unistra.fr/viz-bin/cat/J/A+A/682/A34}}.

% PASA uses footnotes, not endnotes. \endnote in this template will behave like \footnote; and \printendnotes will not output anything.
% \printendnotes

\bibliography{main}

\begin{thebibliography}{}
\expandafter\ifx\csname natexlab\endcsname\relax\def\natexlab#1{#1}\fi

\bibitem[{{Abazajian} {et~al.}(2009){Abazajian}, {Adelman-McCarthy},
  {Ag{\"u}eros}, {Allam}, {Allende Prieto}, {An}, {Anderson}, {Anderson},
  {Annis}, {Bahcall}, {Bailer-Jones}, {Barentine}, {Bassett}, {Becker},
  {Beers}, {Bell}, {Belokurov}, {Berlind}, {Berman}, {Bernardi}, {Bickerton},
  {Bizyaev}, {Blakeslee}, {Blanton}, {Bochanski}, {Boroski}, {Brewington},
  {Brinchmann}, {Brinkmann}, {Brunner}, {Budav{\'a}ri}, {Carey}, {Carliles},
  {Carr}, {Castander}, {Cinabro}, {Connolly}, {Csabai}, {Cunha}, {Czarapata},
  {Davenport}, {de Haas}, {Dilday}, {Doi}, {Eisenstein}, {Evans}, {Evans},
  {Fan}, {Friedman}, {Frieman}, {Fukugita}, {G{\"a}nsicke}, {Gates},
  {Gillespie}, {Gilmore}, {Gonzalez}, {Gonzalez}, {Grebel}, {Gunn},
  {Gy{\"o}ry}, {Hall}, {Harding}, {Harris}, {Harvanek}, {Hawley}, {Hayes},
  {Heckman}, {Hendry}, {Hennessy}, {Hindsley}, {Hoblitt}, {Hogan}, {Hogg},
  {Holtzman}, {Hyde}, {Ichikawa}, {Ichikawa}, {Im}, {Ivezi{\'c}}, {Jester},
  {Jiang}, {Johnson}, {Jorgensen}, {Juri{\'c}}, {Kent}, {Kessler}, {Kleinman},
  {Knapp}, {Konishi}, {Kron}, {Krzesinski}, {Kuropatkin}, {Lampeitl},
  {Lebedeva}, {Lee}, {Lee}, {French Leger}, {L{\'e}pine}, {Li}, {Lima}, {Lin},
  {Long}, {Loomis}, {Loveday}, {Lupton}, {Magnier}, {Malanushenko},
  {Malanushenko}, {Mandelbaum}, {Margon}, {Marriner}, {Mart{\'\i}nez-Delgado},
  {Matsubara}, {McGehee}, {McKay}, {Meiksin}, {Morrison}, {Mullally}, {Munn},
  {Murphy}, {Nash}, {Nebot}, {Neilsen}, {Newberg}, {Newman}, {Nichol},
  {Nicinski}, {Nieto-Santisteban}, {Nitta}, {Okamura}, {Oravetz}, {Ostriker},
  {Owen}, {Padmanabhan}, {Pan}, {Park}, {Pauls}, {Peoples}, {Percival}, {Pier},
  {Pope}, {Pourbaix}, {Price}, {Purger}, {Quinn}, {Raddick}, {Re Fiorentin},
  {Richards}, {Richmond}, {Riess}, {Rix}, {Rockosi}, {Sako}, {Schlegel},
  {Schneider}, {Scholz}, {Schreiber}, {Schwope}, {Seljak}, {Sesar}, {Sheldon},
  {Shimasaku}, {Sibley}, {Simmons}, {Sivarani}, {Allyn Smith}, {Smith},
  {Smol{\v{c}}i{\'c}}, {Snedden}, {Stebbins}, {Steinmetz}, {Stoughton},
  {Strauss}, {SubbaRao}, {Suto}, {Szalay}, {Szapudi}, {Szkody}, {Tanaka},
  {Tegmark}, {Teodoro}, {Thakar}, {Tremonti}, {Tucker}, {Uomoto}, {Vanden
  Berk}, {Vandenberg}, {Vidrih}, {Vogeley}, {Voges}, {Vogt}, {Wadadekar},
  {Watters}, {Weinberg}, {West}, {White}, {Wilhite}, {Wonders}, {Yanny},
  {Yocum}, {York}, {Zehavi}, {Zibetti}, \& {Zucker}}]{2009ApJS..182..543A}
{Abazajian}, K.~N., {Adelman-McCarthy}, J.~K., {Ag{\"u}eros}, M.~A., {et~al.}
  2009, \apjs, 182, 543

\bibitem[{{Abbott} {et~al.}(1986){Abbott}, {Beiging}, {Churchwell}, \&
  {Torres}}]{1986ApJ...303..239A}
{Abbott}, D.~C., {Beiging}, J.~H., {Churchwell}, E., \& {Torres}, A.~V. 1986,
  \apj, 303, 239

\bibitem[{{Andersson} {et~al.}(2022){Andersson}, {Fender}, {Lintott},
  {Williams}, {Driessen}, {Woudt}, {van der Horst}, {Buckley}, {Motta},
  {Rhodes}, {Eisner}, {Osten}, {Vreeswijk}, {Bloemen}, \&
  {Groot}}]{Andersson_Star}
{Andersson}, A., {Fender}, R.~P., {Lintott}, C.~J., {et~al.} 2022, \mnras, 513,
  3482

\bibitem[{{Astropy Collaboration} {et~al.}(2013){Astropy Collaboration},
  {Robitaille}, {Tollerud}, {Greenfield}, {Droettboom}, {Bray}, {Aldcroft},
  {Davis}, {Ginsburg}, {Price-Whelan}, {Kerzendorf}, {Conley}, {Crighton},
  {Barbary}, {Muna}, {Ferguson}, {Grollier}, {Parikh}, {Nair}, {Unther},
  {Deil}, {Woillez}, {Conseil}, {Kramer}, {Turner}, {Singer}, {Fox}, {Weaver},
  {Zabalza}, {Edwards}, {Azalee Bostroem}, {Burke}, {Casey}, {Crawford},
  {Dencheva}, {Ely}, {Jenness}, {Labrie}, {Lim}, {Pierfederici}, {Pontzen},
  {Ptak}, {Refsdal}, {Servillat}, \& {Streicher}}]{2013A&A...558A..33A}
{Astropy Collaboration}, {Robitaille}, T.~P., {Tollerud}, E.~J., {et~al.} 2013,
  \aap, 558, A33

\bibitem[{{Astropy Collaboration} {et~al.}(2018){Astropy Collaboration},
  {Price-Whelan}, {Sip{\H{o}}cz}, {G{\"u}nther}, {Lim}, {Crawford}, {Conseil},
  {Shupe}, {Craig}, {Dencheva}, {Ginsburg}, {Vand erPlas}, {Bradley},
  {P{\'e}rez-Su{\'a}rez}, {de Val-Borro}, {Aldcroft}, {Cruz}, {Robitaille},
  {Tollerud}, {Ardelean}, {Babej}, {Bach}, {Bachetti}, {Bakanov}, {Bamford},
  {Barentsen}, {Barmby}, {Baumbach}, {Berry}, {Biscani}, {Boquien}, {Bostroem},
  {Bouma}, {Brammer}, {Bray}, {Breytenbach}, {Buddelmeijer}, {Burke},
  {Calderone}, {Cano Rodr{\'\i}guez}, {Cara}, {Cardoso}, {Cheedella}, {Copin},
  {Corrales}, {Crichton}, {D'Avella}, {Deil}, {Depagne}, {Dietrich}, {Donath},
  {Droettboom}, {Earl}, {Erben}, {Fabbro}, {Ferreira}, {Finethy}, {Fox},
  {Garrison}, {Gibbons}, {Goldstein}, {Gommers}, {Greco}, {Greenfield},
  {Groener}, {Grollier}, {Hagen}, {Hirst}, {Homeier}, {Horton}, {Hosseinzadeh},
  {Hu}, {Hunkeler}, {Ivezi{\'c}}, {Jain}, {Jenness}, {Kanarek}, {Kendrew},
  {Kern}, {Kerzendorf}, {Khvalko}, {King}, {Kirkby}, {Kulkarni}, {Kumar},
  {Lee}, {Lenz}, {Littlefair}, {Ma}, {Macleod}, {Mastropietro}, {McCully},
  {Montagnac}, {Morris}, {Mueller}, {Mumford}, {Muna}, {Murphy}, {Nelson},
  {Nguyen}, {Ninan}, {N{\"o}the}, {Ogaz}, {Oh}, {Parejko}, {Parley}, {Pascual},
  {Patil}, {Patil}, {Plunkett}, {Prochaska}, {Rastogi}, {Reddy Janga},
  {Sabater}, {Sakurikar}, {Seifert}, {Sherbert}, {Sherwood-Taylor}, {Shih},
  {Sick}, {Silbiger}, {Singanamalla}, {Singer}, {Sladen}, {Sooley},
  {Sornarajah}, {Streicher}, {Teuben}, {Thomas}, {Tremblay}, {Turner},
  {Terr{\'o}n}, {van Kerkwijk}, {de la Vega}, {Watkins}, {Weaver}, {Whitmore},
  {Woillez}, {Zabalza}, \& {Astropy Contributors}}]{2018AJ....156..123A}
{Astropy Collaboration}, {Price-Whelan}, A.~M., {Sip{\H{o}}cz}, B.~M., {et~al.}
  2018, \aj, 156, 123

\bibitem[{{Babusiaux} {et~al.}(2023){Babusiaux}, {Fabricius}, {Khanna},
  {Muraveva}, {Reyl{\'e}}, {Spoto}, {Vallenari}, {Luri}, {Arenou},
  {{\'A}lvarez}, {Anders}, {Antoja}, {Balbinot}, {Barache}, {Bauchet},
  {Bossini}, {Busonero}, {Cantat-Gaudin}, {Carrasco}, {Dafonte}, {Diakit{\'e}},
  {Figueras}, {Garcia-Gutierrez}, {Garofalo}, {Helmi}, {Jim{\'e}nez-Arranz},
  {Jordi}, {Kervella}, {Kostrzewa-Rutkowska}, {Leclerc}, {Licata}, {Manteiga},
  {Masip}, {Mongui{\'o}}, {Ramos}, {Robichon}, {Robin}, {Romero-G{\'o}mez},
  {S{\'a}ez}, {Santove{\~n}a}, {Spina}, {Torralba Elipe}, \&
  {Weiler}}]{GaiaDR3_Validation}
{Babusiaux}, C., {Fabricius}, C., {Khanna}, S., {et~al.} 2023, \aap, 674, A32

\bibitem[{{Bailer-Jones} {et~al.}(2021){Bailer-Jones}, {Rybizki}, {Fouesneau},
  {Demleitner}, \& {Andrae}}]{2021AJ....161..147B}
{Bailer-Jones}, C.~A.~L., {Rybizki}, J., {Fouesneau}, M., {Demleitner}, M., \&
  {Andrae}, R. 2021, \aj, 161, 147

\bibitem[{{Barrett} {et~al.}(2020){Barrett}, {Dieck}, {Beasley}, {Mason}, \&
  {Singh}}]{2020AdSpR..66.1226B}
{Barrett}, P., {Dieck}, C., {Beasley}, A.~J., {Mason}, P.~A., \& {Singh}, K.~P.
  2020, Advances in Space Research, 66, 1226

\bibitem[{{Bastian}(1990)}]{1990SoPh..130..265B}
{Bastian}, T.~S. 1990, \solphys, 130, 265

\bibitem[{{Becker} {et~al.}(1994){Becker}, {White}, \&
  {Helfand}}]{1994ASPC...61..165B}
{Becker}, R.~H., {White}, R.~L., \& {Helfand}, D.~J. 1994, in Astronomical
  Society of the Pacific Conference Series, Vol.~61, Astronomical Data Analysis
  Software and Systems III, ed. D.~R. {Crabtree}, R.~J. {Hanisch}, \&
  J.~{Barnes}, 165

\bibitem[{{Becker} {et~al.}(1995){Becker}, {White}, \&
  {Helfand}}]{1995ApJ...450..559B}
{Becker}, R.~H., {White}, R.~L., \& {Helfand}, D.~J. 1995, \apj, 450, 559

\bibitem[{{Becker} {et~al.}(2001){Becker}, {White}, {Gregg},
  {Laurent-Muehleisen}, {Brotherton}, {Impey}, {Chaffee}, {Richards},
  {Helfand}, {Lacy}, {Courbin}, \& {Proctor}}]{2001ApJS..135..227B}
{Becker}, R.~H., {White}, R.~L., {Gregg}, M.~D., {et~al.} 2001, \apjs, 135, 227

\bibitem[{{Benaglia} {et~al.}(2015){Benaglia}, {Marcote}, {Mold{\'o}n},
  {Nelan}, {De Becker}, {Dougherty}, \& {Koribalski}}]{2015A&A...579A..99B}
{Benaglia}, P., {Marcote}, B., {Mold{\'o}n}, J., {et~al.} 2015, \aap, 579, A99

\bibitem[{{Berger} {et~al.}(2001){Berger}, {Ball}, {Becker}, {Clarke}, {Frail},
  {Fukuda}, {Hoffman}, {Mellon}, {Momjian}, {Murphy}, {Teng}, {Woodruff},
  {Zauderer}, \& {Zavala}}]{2001Natur.410..338B}
{Berger}, E., {Ball}, S., {Becker}, K.~M., {et~al.} 2001, \nat, 410, 338

\bibitem[{{Bernardini} {et~al.}(2013){Bernardini}, {de Martino}, {Mukai},
  {Falanga}, {Andruchow}, {Bonnet-Bidaud}, {Masetti}, {Buitrago}, {Mouchet}, \&
  {Tovmassian}}]{2013MNRAS.435.2822B}
{Bernardini}, F., {de Martino}, D., {Mukai}, K., {et~al.} 2013, \mnras, 435,
  2822

\bibitem[{{Blomme} {et~al.}(2007){Blomme}, {De Becker}, {Runacres}, {van Loo},
  \& {Setia Gunawan}}]{2007A&A...464..701B}
{Blomme}, R., {De Becker}, M., {Runacres}, M.~C., {van Loo}, S., \& {Setia
  Gunawan}, D.~Y.~A. 2007, \aap, 464, 701

\bibitem[{{Callingham} {et~al.}(2019{\natexlab{a}}){Callingham}, {Tuthill},
  {Pope}, {Williams}, {Crowther}, {Edwards}, {Norris}, \&
  {Kedziora-Chudczer}}]{2019NatAs...3...82C}
{Callingham}, J.~R., {Tuthill}, P.~G., {Pope}, B.~J.~S., {et~al.}
  2019{\natexlab{a}}, Nature Astronomy, 3, 82

\bibitem[{{Callingham} {et~al.}(2019{\natexlab{b}}){Callingham}, {Vedantham},
  {Pope}, {Shimwell}, \& {LoTSS Team}}]{2019RNAAS...3...37C}
{Callingham}, J.~R., {Vedantham}, H.~K., {Pope}, B.~J.~S., {Shimwell}, T.~W.,
  \& {LoTSS Team}. 2019{\natexlab{b}}, Research Notes of the American
  Astronomical Society, 3, 37

\bibitem[{{Callingham} {et~al.}(2021){Callingham}, {Vedantham}, {Shimwell},
  {Pope}, {Davis}, {Best}, {Hardcastle}, {R{\"o}ttgering}, {Sabater}, {Tasse},
  {van Weeren}, {Williams}, {Zarka}, {de Gasperin}, \&
  {Drabent}}]{2021NatAs...5.1233C}
{Callingham}, J.~R., {Vedantham}, H.~K., {Shimwell}, T.~W., {et~al.} 2021,
  Nature Astronomy, 5, 1233

\bibitem[{{Callingham} {et~al.}(2023){Callingham}, {Shimwell}, {Vedantham},
  {Bassa}, {O'Sullivan}, {Yiu}, {Bloot}, {Best}, {Hardcastle}, {Haverkorn},
  {Kavanagh}, {Lamy}, {Pope}, {R{\"o}ttgering}, {Schwarz}, {Tasse}, {van
  Weeren}, {White}, {Zarka}, {Bomans}, {Bonafede}, {Bonato}, {Botteon},
  {Bruggen}, {Chy{\.z}y}, {Drabent}, {Emig}, {Gloudemans}, {G{\"u}rkan},
  {Hajduk}, {Hoang}, {Hoeft}, {Iacobelli}, {Kadler}, {Kunert-Bajraszewska},
  {Mingo}, {Morabito}, {Nair}, {P{\'e}rez-Torres}, {Ray}, {Riseley},
  {Rowlinson}, {Shulevski}, {Sweijen}, {Timmerman}, {Vaccari}, \&
  {Zheng}}]{vlotss}
{Callingham}, J.~R., {Shimwell}, T.~W., {Vedantham}, H.~K., {et~al.} 2023,
  \aap, 670, A124

\bibitem[{{Camilo} {et~al.}(2018){Camilo}, {Scholz}, {Serylak}, {Buchner},
  {Merryfield}, {Kaspi}, {Archibald}, {Bailes}, {Jameson}, {van Straten}, \&
  et~al.}]{2018ApJ...856..180C}
{Camilo}, F., {Scholz}, P., {Serylak}, M., {et~al.} 2018, \apj, 856, 180

\bibitem[{{Dougherty} {et~al.}(2005){Dougherty}, {Beasley}, {Claussen},
  {Zauderer}, \& {Bolingbroke}}]{2005ApJ...623..447D}
{Dougherty}, S.~M., {Beasley}, A.~J., {Claussen}, M.~J., {Zauderer}, B.~A., \&
  {Bolingbroke}, N.~J. 2005, \apj, 623, 447

\bibitem[{{Dougherty} \& {Williams}(2000)}]{2000MNRAS.319.1005D}
{Dougherty}, S.~M., \& {Williams}, P.~M. 2000, \mnras, 319, 1005

\bibitem[{{Driessen} {et~al.}(2023){Driessen}, {Heald}, {Duchesne}, {Murphy},
  {Lenc}, {Leung}, \& {Moss}}]{Driessen_ProperMotion}
{Driessen}, L.~N., {Heald}, G., {Duchesne}, S.~W., {et~al.} 2023, \pasa, 40,
  e036

\bibitem[{{Driessen} {et~al.}(2022){Driessen}, {Williams}, {McDonald},
  {Stappers}, {Buckley}, {Fender}, \& {Woudt}}]{Driessen_EXO0408}
{Driessen}, L.~N., {Williams}, D.~R.~A., {McDonald}, I., {et~al.} 2022, \mnras,
  510, 1083

\bibitem[{{Driessen} {et~al.}(2020){Driessen}, {McDonald}, {Buckley}, {Caleb},
  {Kotze}, {Potter}, {Rajwade}, {Rowlinson}, {Stappers}, {Tremou}, {Woudt},
  {Fender}, {Armstrong}, {Groot}, {Heywood}, {Horesh}, {van der Horst},
  {Koerding}, {McBride}, {Miller-Jones}, {Mooley}, \&
  {Wijers}}]{Driessen_Flareyboi}
{Driessen}, L.~N., {McDonald}, I., {Buckley}, D.~A.~H., {et~al.} 2020, \mnras,
  491, 560

\bibitem[{{Duchesne} {et~al.}(2023){Duchesne}, {Thomson}, {Pritchard}, {Lenc},
  {Moss}, {McConnell}, {Wieringa}, {Whiting}, {Wang}, {Wang}, {Rose}, {Raja},
  {Murphy}, {Leung}, {Huynh}, {Hotan}, {Hodgson}, \&
  {Heald}}]{2023PASA...40...34D}
{Duchesne}, S.~W., {Thomson}, A.~J.~M., {Pritchard}, J., {et~al.} 2023, \pasa,
  40, e034

\bibitem[{{Dulk}(1985)}]{1985ARA&A..23..169D}
{Dulk}, G.~A. 1985, \araa, 23, 169

\bibitem[{{Duncan} \& {White}(2002)}]{2002MNRAS.330...63D}
{Duncan}, R.~A., \& {White}, S.~M. 2002, \mnras, 330, 63

\bibitem[{{Fabricius} {et~al.}(2021){Fabricius}, {Luri}, {Arenou}, {Babusiaux},
  {Helmi}, {Muraveva}, {Reyl{\'e}}, {Spoto}, {Vallenari}, {Antoja}, {Balbinot},
  {Barache}, {Bauchet}, {Bragaglia}, {Busonero}, {Cantat-Gaudin}, {Carrasco},
  {Diakit{\'e}}, {Fabrizio}, {Figueras}, {Garcia-Gutierrez}, {Garofalo},
  {Jordi}, {Kervella}, {Khanna}, {Leclerc}, {Licata}, {Lambert}, {Marrese},
  {Masip}, {Ramos}, {Robichon}, {Robin}, {Romero-G{\'o}mez}, {Rubele}, \&
  {Weiler}}]{GaiaEDR3_Validation}
{Fabricius}, C., {Luri}, X., {Arenou}, F., {et~al.} 2021, \aap, 649, A5

\bibitem[{{Fekel}(1996)}]{1996AJ....112..269F}
{Fekel}, F.~C. 1996, \aj, 112, 269

\bibitem[{{Fender} {et~al.}(2016){Fender}, {Woudt}, {Corbel}, {Coriat},
  {Daigne}, {Falcke}, {Girard}, {Heywood}, {Horesh}, {Horrell}, {Jonker},
  {Joseph}, {Kamble}, {Knigge}, {K{\"o}rding}, {Kotze}, {Kouveliotou}, {Lynch},
  {Maccarone}, {Meintjes}, {Migliari}, {Murphy}, {Nagayama}, {Nelemans},
  {Nicholson}, {O'Brien}, {Oodendaal}, {Oozeer}, {Osborne}, {P{\'e}rez-Torres},
  {Ratcliffe}, {Ribeiro}, {Rol}, {Rushton}, {Scaife}, {Schurch}, {Sivakoff},
  {Staley}, {Steeghs}, {Stewart}, {Swinbank}, {Vergani}, {Warner}, {Wiersema},
  {Armstrong}, {Groot}, {McBride}, {Miller-Jones}, {Mooley}, {Stappers},
  {Wijers}, {Bietenholz}, {Blyth}, {B{\"o}ttcher}, {Buckley}, {Charles},
  {Chomiuk}, {Coppejans}, {de Blok}, {van der Heyden}, {van der Horst}, \& {van
  Soelen}}]{TKT_description}
{Fender}, R., {Woudt}, P.~A., {Corbel}, S., {et~al.} 2016, in MeerKAT Science:
  On the Pathway to the SKA, 13

\bibitem[{{Flesch}(2023)}]{milliquas}
{Flesch}, E.~W. 2023, The Open Journal of Astrophysics, 6, 49

\bibitem[{{Freund} {et~al.}(2022){Freund}, {Czesla}, {Robrade}, {Schneider}, \&
  {Schmitt}}]{2022A&A...664A.105F}
{Freund}, S., {Czesla}, S., {Robrade}, J., {Schneider}, P.~C., \& {Schmitt},
  J.~H.~M.~M. 2022, \aap, 664, A105

\bibitem[{{Freund} {et~al.}(2018){Freund}, {Robrade}, {Schneider}, \&
  {Schmitt}}]{2018A&A...614A.125F}
{Freund}, S., {Robrade}, J., {Schneider}, P.~C., \& {Schmitt}, J.~H.~M.~M.
  2018, \aap, 614, A125

\bibitem[{{Freund} {et~al.}(2024){Freund}, {Czesla}, {Predehl}, {Robrade},
  {Salvato}, {Schneider}, {Starck}, {Wolf}, \& {Schmitt}}]{2024arXiv240117282F}
{Freund}, S., {Czesla}, S., {Predehl}, P., {et~al.} 2024, arXiv e-prints,
  arXiv:2401.17282

\bibitem[{{Gaia Collaboration} {et~al.}(2016{\natexlab{a}}){Gaia
  Collaboration}, {Brown}, {Vallenari}, {Prusti}, {de Bruijne}, {Mignard},
  {Drimmel}, {Babusiaux}, {Bailer-Jones}, {Bastian}, \& et~al.}]{GaiaDR1}
{Gaia Collaboration}, {Brown}, A.~G.~A., {Vallenari}, A., {et~al.}
  2016{\natexlab{a}}, \aap, 595, A2

\bibitem[{{Gaia Collaboration} {et~al.}(2016{\natexlab{b}}){Gaia
  Collaboration}, {Prusti}, {de Bruijne}, {Brown}, {Vallenari}, {Babusiaux},
  {Bailer-Jones}, {Bastian}, {Biermann}, {Evans}, \& et~al.}]{GaiaMission}
{Gaia Collaboration}, {Prusti}, T., {de Bruijne}, J.~H.~J., {et~al.}
  2016{\natexlab{b}}, \aap, 595, A1

\bibitem[{{Gaia Collaboration} {et~al.}(2018){Gaia Collaboration}, {Brown},
  {Vallenari}, {Prusti}, {de Bruijne}, {Babusiaux}, {Bailer-Jones}, {Biermann},
  {Evans}, {Eyer}, \& et~al.}]{GaiaDR2}
{Gaia Collaboration}, {Brown}, A.~G.~A., {Vallenari}, A., {et~al.} 2018, \aap,
  616, A1

\bibitem[{{Gaia Collaboration} {et~al.}(2021{\natexlab{a}}){Gaia
  Collaboration}, {Brown}, {Vallenari}, {Prusti}, {de Bruijne}, {Babusiaux},
  {Biermann}, {Creevey}, {Evans}, {Eyer}, \& et~al.}]{GaiaEDR3_Summary}
---. 2021{\natexlab{a}}, \aap, 649, A1

\bibitem[{{Gaia Collaboration} {et~al.}(2021{\natexlab{b}}){Gaia
  Collaboration}, {Smart}, {Sarro}, {Rybizki}, {Reyl{\'e}}, {Robin}, {Hambly},
  {Abbas}, {Barstow}, {de Bruijne}, \& et~al.}]{2021A&A...649A...6G}
{Gaia Collaboration}, {Smart}, R.~L., {Sarro}, L.~M., {et~al.}
  2021{\natexlab{b}}, \aap, 649, A6

\bibitem[{{Gaia Collaboration} {et~al.}(2023){Gaia Collaboration}, {Vallenari},
  {Brown}, {Prusti}, {de Bruijne}, {Arenou}, {Babusiaux}, {Biermann},
  {Creevey}, {Ducourant}, \& et~al.}]{GaiaDR3}
{Gaia Collaboration}, {Vallenari}, A., {Brown}, A.~G.~A., {et~al.} 2023, \aap,
  674, A1

\bibitem[{{Golovin} {et~al.}(2023){Golovin}, {Reffert}, {Just}, {Jordan},
  {Vani}, \& {Jahrei{\ss}}}]{2023A&A...670A..19G}
{Golovin}, A., {Reffert}, S., {Just}, A., {et~al.} 2023, \aap, 670, A19

\bibitem[{{Gordon} {et~al.}(2021){Gordon}, {Boyce}, {O'Dea}, {Rudnick},
  {Andernach}, {Vantyghem}, {Baum}, {Bui}, {Dionyssiou}, {Safi-Harb}, \&
  {Sander}}]{2021ApJS..255...30G}
{Gordon}, Y.~A., {Boyce}, M.~M., {O'Dea}, C.~P., {et~al.} 2021, \apjs, 255, 30

\bibitem[{{Guedel} \& {Benz}(1993)}]{1993ApJ...405L..63G}
{Guedel}, M., \& {Benz}, A.~O. 1993, \apjl, 405, L63

\bibitem[{{Guns} {et~al.}(2021){Guns}, {Foster}, {Daley}, {Rahlin},
  {Whitehorn}, {Ade}, {Ahmed}, {Anderes}, {Anderson}, {Archipley}, {Avva},
  {Aylor}, {Balkenhol}, {Barry}, {Basu Thakur}, {Benabed}, {Bender}, {Benson},
  {Bianchini}, {Bleem}, {Bouchet}, {Bryant}, {Byrum}, {Carlstrom}, {Carter},
  {Cecil}, {Chang}, {Chaubal}, {Chen}, {Cho}, {Chou}, {Cliche}, {Crawford},
  {Cukierman}, {de Haan}, {Denison}, {Dibert}, {Ding}, {Dobbs}, {Dutcher},
  {Everett}, {Feng}, {Ferguson}, {Fu}, {Galli}, {Gambrel}, {Gardner},
  {Goeckner-Wald}, {Gualtieri}, {Gupta}, {Guyser}, {Halverson},
  {Harke-Hosemann}, {Harrington}, {Henning}, {Hilton}, {Hivon}, {Holder},
  {Holzapfel}, {Hood}, {Howe}, {Huang}, {Irwin}, {Jeong}, {Jonas}, {Jones},
  {Khaire}, {Knox}, {Kofman}, {Korman}, {Kubik}, {Kuhlmann}, {Kuo}, {Lee},
  {Leitch}, {Lowitz}, {Lu}, {Marrone}, {Meyer}, {Michalik}, {Millea},
  {Montgomery}, {Nadolski}, {Natoli}, {Nguyen}, {Noble}, {Novosad}, {Omori},
  {Padin}, {Pan}, {Paschos}, {Pearson}, {Phadke}, {Posada}, {Prabhu}, {Quan},
  {Reichardt}, {Riebel}, {Riedel}, {Rouble}, {Ruhl}, {Sayre}, {Schiappucci},
  {Shirokoff}, {Smecher}, {Sobrin}, {Stark}, {Stephen}, {Story}, {Suzuki},
  {Thompson}, {Thorne}, {Tucker}, {Umilta}, {Vale}, {Vieira}, {Wang}, {Wu},
  {Yefremenko}, {Yoon}, {Young}, \& {Zhang}}]{2021ApJ...916...98G}
{Guns}, S., {Foster}, A., {Daley}, C., {et~al.} 2021, \apj, 916, 98

\bibitem[{{Haisch} {et~al.}(1978){Haisch}, {Linsky}, {Slee}, {Hearn}, {Walker},
  {Rydgren}, \& {Nicolson}}]{1978ApJ...225L..35H}
{Haisch}, B.~M., {Linsky}, J.~L., {Slee}, O.~B., {et~al.} 1978, \apjl, 225, L35

\bibitem[{{Hale} {et~al.}(2021){Hale}, {McConnell}, {Thomson}, {Lenc}, {Heald},
  {Hotan}, {Leung}, {Moss}, {Murphy}, {Pritchard}, {Sadler}, {Stewart}, \&
  {Whiting}}]{2021PASA...38...58H}
{Hale}, C.~L., {McConnell}, D., {Thomson}, A.~J.~M., {et~al.} 2021, \pasa, 38,
  e058

\bibitem[{{Helfand} {et~al.}(1999){Helfand}, {Schnee}, {Becker}, {White}, \&
  {McMahon}}]{1999AJ....117.1568H}
{Helfand}, D.~J., {Schnee}, S., {Becker}, R.~H., {White}, R.~L., \& {McMahon},
  R.~G. 1999, \aj, 117, 1568

\bibitem[{{Hewish} {et~al.}(1968){Hewish}, {Bell}, {Pilkington}, {Scott}, \&
  {Collins}}]{1968Natur.217..709H}
{Hewish}, A., {Bell}, S.~J., {Pilkington}, J.~D.~H., {Scott}, P.~F., \&
  {Collins}, R.~A. 1968, \nat, 217, 709

\bibitem[{{Hindson} {et~al.}(2012){Hindson}, {Thompson}, {Urquhart}, {Faimali},
  {Clark}, \& {Davies}}]{2012MNRAS.421.3418H}
{Hindson}, L., {Thompson}, M.~A., {Urquhart}, J.~S., {et~al.} 2012, \mnras,
  421, 3418

\bibitem[{{H{\o}g} {et~al.}(2000{\natexlab{a}}){H{\o}g}, {Fabricius},
  {Makarov}, {Bastian}, {Schwekendiek}, {Wicenec}, {Urban}, {Corbin}, \&
  {Wycoff}}]{TYCHO_2_Validation}
{H{\o}g}, E., {Fabricius}, C., {Makarov}, V.~V., {et~al.} 2000{\natexlab{a}},
  \aap, 357, 367

\bibitem[{{H{\o}g} {et~al.}(2000{\natexlab{b}}){H{\o}g}, {Fabricius},
  {Makarov}, {Urban}, {Corbin}, {Wycoff}, {Bastian}, {Schwekendiek}, \&
  {Wicenec}}]{TYCHO_2}
---. 2000{\natexlab{b}}, \aap, 355, L27

\bibitem[{{Hotan} {et~al.}(2021){Hotan}, {Bunton}, {Chippendale}, {Whiting},
  {Tuthill}, {Moss}, {McConnell}, {Amy}, {Huynh}, {Allison}, {Anderson},
  {Bannister}, {Bastholm}, {Beresford}, {Bock}, {Bolton}, {Chapman}, {Chow},
  {Collier}, {Cooray}, {Cornwell}, {Diamond}, {Edwards}, {Feain}, {Franzen},
  {George}, {Gupta}, {Hampson}, {Harvey-Smith}, {Hayman}, {Heywood}, {Jacka},
  {Jackson}, {Jackson}, {Jeganathan}, {Johnston}, {Kesteven}, {Kleiner},
  {Koribalski}, {Lee-Waddell}, {Lenc}, {Lensson}, {Mackay}, {Mahony},
  {McClure-Griffiths}, {McConigley}, {Mirtschin}, {Ng}, {Norris}, {Pearce},
  {Phillips}, {Pilawa}, {Raja}, {Reynolds}, {Roberts}, {Roxby}, {Sadler},
  {Shields}, {Schinckel}, {Serra}, {Shaw}, {Sweetnam}, {Troup}, {Tzioumis},
  {Voronkov}, \& {Westmeier}}]{2021PASA...38....9H}
{Hotan}, A.~W., {Bunton}, J.~D., {Chippendale}, A.~P., {et~al.} 2021, \pasa,
  38, e009

\bibitem[{{Jones} {et~al.}(2009){Jones}, {Read}, {Saunders}, {Colless},
  {Jarrett}, {Parker}, {Fairall}, {Mauch}, {Sadler}, {Watson}, {Burton},
  {Campbell}, {Cass}, {Croom}, {Dawe}, {Fiegert}, {Frankcombe}, {Hartley},
  {Huchra}, {James}, {Kirby}, {Lahav}, {Lucey}, {Mamon}, {Moore}, {Peterson},
  {Prior}, {Proust}, {Russell}, {Safouris}, {Wakamatsu}, {Westra}, \&
  {Williams}}]{2009MNRAS.399..683J}
{Jones}, D.~H., {Read}, M.~A., {Saunders}, W., {et~al.} 2009, \mnras, 399, 683

\bibitem[{{Jones} {et~al.}(2005){Jones}, {Saunders}, {Colless}, {Read},
  {Parker}, {Watson}, \& {Campbell}}]{2005ASPC..329...11J}
{Jones}, H., {Saunders}, W., {Colless}, M., {et~al.} 2005, in Astronomical
  Society of the Pacific Conference Series, Vol. 329, Nearby Large-Scale
  Structures and the Zone of Avoidance, ed. A.~P. {Fairall} \& P.~A. {Woudt},
  11

\bibitem[{{Kao} {et~al.}(2018){Kao}, {Hallinan}, {Pineda}, {Stevenson}, \&
  {Burgasser}}]{2018ApJS..237...25K}
{Kao}, M.~M., {Hallinan}, G., {Pineda}, J.~S., {Stevenson}, D., \& {Burgasser},
  A. 2018, \apjs, 237, 25

\bibitem[{{Kawai} {et~al.}(2022){Kawai}, {Tsuboi}, {Iwakiri}, {Maeda},
  {Katsuda}, {Sasaki}, {Kohara}, \& {MAXI TEAM}}]{2022PASJ...74..477K}
{Kawai}, H., {Tsuboi}, Y., {Iwakiri}, W.~B., {et~al.} 2022, \pasj, 74, 477

\bibitem[{{Kimball} {et~al.}(2009){Kimball}, {Knapp}, {Ivezi{\'c}}, {West},
  {Bochanski}, {Plotkin}, \& {Gordon}}]{2009ApJ...701..535K}
{Kimball}, A.~E., {Knapp}, G.~R., {Ivezi{\'c}}, {\v{Z}}., {et~al.} 2009, \apj,
  701, 535

\bibitem[{{Lacy} {et~al.}(2020){Lacy}, {Baum}, {Chandler}, {Chatterjee},
  {Clarke}, {Deustua}, {English}, {Farnes}, {Gaensler}, {Gugliucci},
  {Hallinan}, {Kent}, {Kimball}, {Law}, {Lazio}, {Marvil}, {Mao}, {Medlin},
  {Mooley}, {Murphy}, {Myers}, {Osten}, {Richards}, {Rosolowsky}, {Rudnick},
  {Schinzel}, {Sivakoff}, {Sjouwerman}, {Taylor}, {White}, {Wrobel},
  {Andernach}, {Beasley}, {Berger}, {Bhatnager}, {Birkinshaw}, {Bower},
  {Brandt}, {Brown}, {Burke-Spolaor}, {Butler}, {Comerford}, {Demorest}, {Fu},
  {Giacintucci}, {Golap}, {G{\"u}th}, {Hales}, {Hiriart}, {Hodge}, {Horesh},
  {Ivezi{\'c}}, {Jarvis}, {Kamble}, {Kassim}, {Liu}, {Loinard}, {Lyons},
  {Masters}, {Mezcua}, {Moellenbrock}, {Mroczkowski}, {Nyland}, {O'Dea},
  {O'Sullivan}, {Peters}, {Radford}, {Rao}, {Robnett}, {Salcido}, {Shen},
  {Sobotka}, {Witz}, {Vaccari}, {van Weeren}, {Vargas}, {Williams}, \&
  {Yoon}}]{2020PASP..132c5001L}
{Lacy}, M., {Baum}, S.~A., {Chandler}, C.~J., {et~al.} 2020, \pasp, 132, 035001

\bibitem[{{Lenc} {et~al.}(2018){Lenc}, {Murphy}, {Lynch}, {Kaplan}, \&
  {Zhang}}]{2018MNRAS.478.2835L}
{Lenc}, E., {Murphy}, T., {Lynch}, C.~R., {Kaplan}, D.~L., \& {Zhang}, S.~N.
  2018, \mnras, 478, 2835

\bibitem[{{Lovell} {et~al.}(1963){Lovell}, {Whipple}, \&
  {Solomon}}]{1963Natur.198}
{Lovell}, B., {Whipple}, F.~L., \& {Solomon}, L.~H. 1963, \nat, 198, 228

\bibitem[{{Masetti} {et~al.}(2010){Masetti}, {Parisi}, {Palazzi},
  {Jim{\'e}nez-Bail{\'o}n}, {Chavushyan}, {Bassani}, {Bazzano}, {Bird}, {Dean},
  {Charles}, {Galaz}, {Landi}, {Malizia}, {Mason}, {McBride}, {Minniti},
  {Morelli}, {Schiavone}, {Stephen}, \& {Ubertini}}]{2010A&A...519A..96M}
{Masetti}, N., {Parisi}, P., {Palazzi}, E., {et~al.} 2010, \aap, 519, A96

\bibitem[{{Matthews}(2019)}]{2019PASP..131a6001M}
{Matthews}, L.~D. 2019, \pasp, 131, 016001

\bibitem[{{McConnell} {et~al.}(2020){McConnell}, {Hale}, {Lenc}, {Banfield},
  {Heald}, {Hotan}, {Leung}, {Moss}, {Murphy}, {O'Brien}, {Pritchard}, {Raja},
  {Sadler}, {Stewart}, {Thomson}, {Whiting}, {Allison}, {Amy}, {Anderson},
  {Ball}, {Bannister}, {Bell}, {Bock}, {Bolton}, {Bunton}, {Chippendale},
  {Collier}, {Cooray}, {Cornwell}, {Diamond}, {Edwards}, {Gupta}, {Hayman},
  {Heywood}, {Jackson}, {Koribalski}, {Lee-Waddell}, {McClure-Griffiths}, {Ng},
  {Norris}, {Phillips}, {Reynolds}, {Roxby}, {Schinckel}, {Shields},
  {Tremblay}, {Tzioumis}, {Voronkov}, \& {Westmeier}}]{2020PASA...37...48M}
{McConnell}, D., {Hale}, C.~L., {Lenc}, E., {et~al.} 2020, \pasa, 37, e048

\bibitem[{{Merloni} {et~al.}(2024){Merloni}, {Lamer}, {Liu}, {Ramos-Ceja},
  {Brunner}, {Bulbul}, {Dennerl}, {Doroshenko}, {Freyberg}, {Friedrich},
  {Gatuzz}, {Georgakakis}, {Haberl}, {Igo}, {Kreykenbohm}, {Liu}, {Maitra},
  {Malyali}, {Mayer}, {Nandra}, {Predehl}, {Robrade}, {Salvato}, {Sanders},
  {Stewart}, {Tub{\'\i}n-Arenas}, {Weber}, {Wilms}, {Arcodia}, {Artis},
  {Aschersleben}, {Avakyan}, {Aydar}, {Bahar}, {Balzer}, {Becker}, {Berger},
  {Boller}, {Bornemann}, {Br{\"u}ggen}, {Brusa}, {Buchner}, {Burwitz},
  {Camilloni}, {Clerc}, {Comparat}, {Coutinho}, {Czesla}, {Dannhauer},
  {Dauner}, {Dauser}, {Dietl}, {Dolag}, {Dwelly}, {Egg}, {Ehl}, {Freund},
  {Friedrich}, {Gaida}, {Garrel}, {Ghirardini}, {Gokus}, {Gr{\"u}nwald},
  {Grandis}, {Grotova}, {Gruen}, {Gueguen}, {H{\"a}mmerich}, {Hamaus},
  {Hasinger}, {Haubner}, {Homan}, {Ider Chitham}, {Joseph}, {Joyce},
  {K{\"o}nig}, {Kaltenbrunner}, {Khokhriakova}, {Kink}, {Kirsch}, {Kluge},
  {Knies}, {Krippendorf}, {Krumpe}, {Kurpas}, {Li}, {Liu}, {Locatelli},
  {Lorenz}, {M{\"u}ller}, {Magaudda}, {Mannes}, {McCall}, {Meidinger},
  {Michailidis}, {Migkas}, {Mu{\~n}oz-Giraldo}, {Musiimenta}, {Nguyen-Dang},
  {Ni}, {Olechowska}, {Ota}, {Pacaud}, {Pasini}, {Perinati}, {Pires},
  {Pommranz}, {Ponti}, {Poppenhaeger}, {P{\"u}hlhofer}, {Rau}, {Reh},
  {Reiprich}, {Roster}, {Saeedi}, {Santangelo}, {Sasaki}, {Schmitt},
  {Schneider}, {Schrabback}, {Schuster}, {Schwope}, {Seppi}, {Serim},
  {Shreeram}, {Sokolova-Lapa}, {Starck}, {Stelzer}, {Stierhof}, {Suleimanov},
  {Tenzer}, {Traulsen}, {Tr{\"u}mper}, {Tsuge}, {Urrutia}, {Veronica},
  {Waddell}, {Willer}, {Wolf}, {Yeung}, {Zainab}, {Zangrandi}, {Zhang},
  {Zhang}, \& {Zheng}}]{eROSITA_DR1}
{Merloni}, A., {Lamer}, G., {Liu}, T., {et~al.} 2024, \aap, 682, A34

\bibitem[{{Murphy} {et~al.}(2013){Murphy}, {Chatterjee}, {Kaplan}, {Banyer},
  {Bell}, {Bignall}, {Bower}, {Cameron}, {Coward}, {Cordes}, {Croft}, {Curran},
  {Djorgovski}, {Farrell}, {Frail}, {Gaensler}, {Galloway}, {Gendre}, {Green},
  {Hancock}, {Johnston}, {Kamble}, {Law}, {Lazio}, {Lo}, {Macquart}, {Rea},
  {Rebbapragada}, {Reynolds}, {Ryder}, {Schmidt}, {Soria}, {Stairs}, {Tingay},
  {Torkelsson}, {Wagstaff}, {Walker}, {Wayth}, \& {Williams}}]{VAST}
{Murphy}, T., {Chatterjee}, S., {Kaplan}, D.~L., {et~al.} 2013, \pasa, 30, e006

\bibitem[{{Murphy} {et~al.}(2021){Murphy}, {Kaplan}, {Stewart}, {O'Brien},
  {Lenc}, {Pintaldi}, {Pritchard}, {Dobie}, {Fox}, {Leung}, {An}, {Bell},
  {Broderick}, {Chatterjee}, {Dai}, {d'Antonio}, {Doyle}, {Gaensler}, {Heald},
  {Horesh}, {Jones}, {McConnell}, {Moss}, {Raja}, {Ramsay}, {Ryder}, {Sadler},
  {Sivakoff}, {Wang}, {Wang}, {Wheatland}, {Whiting}, {Allison}, {Anderson},
  {Ball}, {Bannister}, {Bock}, {Bolton}, {Bunton}, {Chekkala}, {Chippendale},
  {Cooray}, {Gupta}, {Hayman}, {Jeganathan}, {Koribalski}, {Lee-Waddell},
  {Mahony}, {Marvil}, {McClure-Griffiths}, {Mirtschin}, {Ng}, {Pearce},
  {Phillips}, \& {Voronkov}}]{Murphy2021}
{Murphy}, T., {Kaplan}, D.~L., {Stewart}, A.~J., {et~al.} 2021, \pasa, 38, e054

\bibitem[{{Ochsenbein} {et~al.}(2000){Ochsenbein}, {Bauer}, \&
  {Marcout}}]{2000A&AS..143...23O}
{Ochsenbein}, F., {Bauer}, P., \& {Marcout}, J. 2000, \aaps, 143, 23

\bibitem[{{O'Sullivan} {et~al.}(2013){O'Sullivan}, {McClure-Griffiths},
  {Feain}, {Gaensler}, \& {Sault}}]{2013MNRAS.435..311O}
{O'Sullivan}, S.~P., {McClure-Griffiths}, N.~M., {Feain}, I.~J., {Gaensler},
  B.~M., \& {Sault}, R.~J. 2013, \mnras, 435, 311

\bibitem[{{Pedersen} {et~al.}(2019){Pedersen}, {Chowdhury}, {Johnston},
  {Bowman}, {Aerts}, {Handler}, {De Cat}, {Neiner}, {David-Uraz}, {Buzasi},
  {Tkachenko}, {Sim{\'o}n-D{\'\i}az}, {Moravveji}, {Sikora}, {Mirouh},
  {Lovekin}, {Cantiello}, {Daszy{\'n}ska-Daszkiewicz}, {Pigulski},
  {Vanderspek}, \& {Ricker}}]{2019ApJ...872L...9P}
{Pedersen}, M.~G., {Chowdhury}, S., {Johnston}, C., {et~al.} 2019, \apjl, 872,
  L9

\bibitem[{{Perley} {et~al.}(2011){Perley}, {Chandler}, {Butler}, \&
  {Wrobel}}]{perley2011}
{Perley}, R.~A., {Chandler}, C.~J., {Butler}, B.~J., \& {Wrobel}, J.~M. 2011,
  \apjl, 739, L1

\bibitem[{{Perryman} {et~al.}(1997){Perryman}, {Lindegren}, {Kovalevsky},
  {Hoeg}, {Bastian}, {Bernacca}, {Cr{\'e}z{\'e}}, {Donati}, {Grenon},
  {Grewing}, {van Leeuwen}, {van der Marel}, {Mignard}, {Murray}, {Le Poole},
  {Schrijver}, {Turon}, {Arenou}, {Froeschl{\'e}}, \& {Petersen}}]{hipparcos}
{Perryman}, M.~A.~C., {Lindegren}, L., {Kovalevsky}, J., {et~al.} 1997, \aap,
  323, L49

\bibitem[{{Predehl} {et~al.}(2021){Predehl}, {Andritschke}, {Arefiev},
  {Babyshkin}, {Batanov}, {Becker}, {B{\"o}hringer}, {Bogomolov}, {Boller},
  {Borm}, {Bornemann}, {Br{\"a}uninger}, {Br{\"u}ggen}, {Brunner}, {Brusa},
  {Bulbul}, {Buntov}, {Burwitz}, {Burkert}, {Clerc}, {Churazov}, {Coutinho},
  {Dauser}, {Dennerl}, {Doroshenko}, {Eder}, {Emberger}, {Eraerds},
  {Finoguenov}, {Freyberg}, {Friedrich}, {Friedrich}, {F{\"u}rmetz},
  {Georgakakis}, {Gilfanov}, {Granato}, {Grossberger}, {Gueguen}, {Gureev},
  {Haberl}, {H{\"a}lker}, {Hartner}, {Hasinger}, {Huber}, {Ji}, {Kienlin},
  {Kink}, {Korotkov}, {Kreykenbohm}, {Lamer}, {Lomakin}, {Lapshov}, {Liu},
  {Maitra}, {Meidinger}, {Menz}, {Merloni}, {Mernik}, {Mican}, {Mohr},
  {M{\"u}ller}, {Nandra}, {Nazarov}, {Pacaud}, {Pavlinsky}, {Perinati},
  {Pfeffermann}, {Pietschner}, {Ramos-Ceja}, {Rau}, {Reiffers}, {Reiprich},
  {Robrade}, {Salvato}, {Sanders}, {Santangelo}, {Sasaki}, {Scheuerle},
  {Schmid}, {Schmitt}, {Schwope}, {Shirshakov}, {Steinmetz}, {Stewart},
  {Str{\"u}der}, {Sunyaev}, {Tenzer}, {Tiedemann}, {Tr{\"u}mper}, {Voron},
  {Weber}, {Wilms}, \& {Yaroshenko}}]{2021A&A...647A...1P}
{Predehl}, P., {Andritschke}, R., {Arefiev}, V., {et~al.} 2021, \aap, 647, A1

\bibitem[{{Pritchard} {et~al.}(2024){Pritchard}, {Murphy}, {Heald},
  {Wheatland}, {Kaplan}, {Lenc}, {O'Brien}, \& {Wang}}]{2024MNRAS.tmp..161P}
{Pritchard}, J., {Murphy}, T., {Heald}, G., {et~al.} 2024, \mnras, 529, 1258

\bibitem[{{Pritchard} {et~al.}(2021){Pritchard}, {Murphy}, {Zic}, {Lynch},
  {Heald}, {Kaplan}, {Anderson}, {Banfield}, {Hale}, {Hotan}, {Lenc}, {Leung},
  {McConnell}, {Moss}, {Raja}, {Stewart}, \& {Whiting}}]{2021MNRAS.502.5438P}
{Pritchard}, J., {Murphy}, T., {Zic}, A., {et~al.} 2021, \mnras, 502, 5438

\bibitem[{{Robitaille} \& {Bressert}(2012)}]{2012ascl.soft08017R}
{Robitaille}, T., \& {Bressert}, E. 2012, {APLpy: Astronomical Plotting Library
  in Python}, Astrophysics Source Code Library, ascl:1208.017

\bibitem[{{Rodet} {et~al.}(2018){Rodet}, {Bonnefoy}, {Durkan}, {Beust},
  {Lagrange}, {Schlieder}, {Janson}, {Grandjean}, {Chauvin}, {Messina},
  {Maire}, {Brandner}, {Girard}, {Delorme}, {Biller}, {Bergfors}, {Lacour},
  {Feldt}, {Henning}, {Boccaletti}, {Le Bouquin}, {Berger}, {Monin}, {Udry},
  {Peretti}, {Segransan}, {Allard}, {Homeier}, {Vigan}, {Langlois},
  {Hagelberg}, {Menard}, {Bazzon}, {Beuzit}, {Delboulb{\'e}}, {Desidera},
  {Gratton}, {Lannier}, {Ligi}, {Maurel}, {Mesa}, {Meyer}, {Pavlov}, {Ramos},
  {Rigal}, {Roelfsema}, {Salter}, {Samland}, {Schmidt}, {Stadler}, \&
  {Weber}}]{2018A&A...618A..23R}
{Rodet}, L., {Bonnefoy}, M., {Durkan}, S., {et~al.} 2018, \aap, 618, A23

\bibitem[{{Rosslowe} \& {Crowther}(2015)}]{2015MNRAS.447.2322R}
{Rosslowe}, C.~K., \& {Crowther}, P.~A. 2015, \mnras, 447, 2322

\bibitem[{{Route} \& {Wolszczan}(2012)}]{2012ApJ...747L..22R}
{Route}, M., \& {Wolszczan}, A. 2012, \apjl, 747, L22

\bibitem[{{Saikia} \& {Salter}(1988)}]{1988ARA&A..26...93S}
{Saikia}, D.~J., \& {Salter}, C.~J. 1988, \araa, 26, 93

\bibitem[{{Samus'} {et~al.}(2017){Samus'}, {Kazarovets}, {Durlevich},
  {Kireeva}, \& {Pastukhova}}]{GCVS}
{Samus'}, N.~N., {Kazarovets}, E.~V., {Durlevich}, O.~V., {Kireeva}, N.~N., \&
  {Pastukhova}, E.~N. 2017, Astronomy Reports, 61, 80

\bibitem[{{Saxton} {et~al.}(2008){Saxton}, {Read}, {Esquej}, {Freyberg},
  {Altieri}, \& {Bermejo}}]{2008A&A...480..611S}
{Saxton}, R.~D., {Read}, A.~M., {Esquej}, P., {et~al.} 2008, \aap, 480, 611

\bibitem[{{Seaquist}(1976)}]{1976ApJ...203L..35S}
{Seaquist}, E.~R. 1976, \apjl, 203, L35

\bibitem[{{Shimwell} {et~al.}(2017){Shimwell}, {R{\"o}ttgering}, {Best},
  {Williams}, {Dijkema}, {de Gasperin}, {Hardcastle}, {Heald}, {Hoang},
  {Horneffer}, {Intema}, {Mahony}, {Mandal}, {Mechev}, {Morabito}, {Oonk},
  {Rafferty}, {Retana-Montenegro}, {Sabater}, {Tasse}, {van Weeren},
  {Br{\"u}ggen}, {Brunetti}, {Chy{\.z}y}, {Conway}, {Haverkorn}, {Jackson},
  {Jarvis}, {McKean}, {Miley}, {Morganti}, {White}, {Wise}, {van Bemmel},
  {Beck}, {Brienza}, {Bonafede}, {Calistro Rivera}, {Cassano}, {Clarke},
  {Cseh}, {Deller}, {Drabent}, {van Driel}, {Engels}, {Falcke}, {Ferrari},
  {Fr{\"o}hlich}, {Garrett}, {Harwood}, {Heesen}, {Hoeft}, {Horellou},
  {Israel}, {Kapi{\'n}ska}, {Kunert-Bajraszewska}, {McKay}, {Mohan},
  {Orr{\'u}}, {Pizzo}, {Prandoni}, {Schwarz}, {Shulevski}, {Sipior}, {Smith},
  {Sridhar}, {Steinmetz}, {Stroe}, {Varenius}, {van der Werf}, {Zensus}, \&
  {Zwart}}]{2017A&A...598A.104S}
{Shimwell}, T.~W., {R{\"o}ttgering}, H.~J.~A., {Best}, P.~N., {et~al.} 2017,
  \aap, 598, A104

\bibitem[{{Shimwell} {et~al.}(2022){Shimwell}, {Hardcastle}, {Tasse}, {Best},
  {R{\"o}ttgering}, {Williams}, {Botteon}, {Drabent}, {Mechev}, {Shulevski},
  {van Weeren}, {Bester}, {Br{\"u}ggen}, {Brunetti}, {Callingham}, {Chy{\.z}y},
  {Conway}, {Dijkema}, {Duncan}, {de Gasperin}, {Hale}, {Haverkorn}, {Hugo},
  {Jackson}, {Mevius}, {Miley}, {Morabito}, {Morganti}, {Offringa}, {Oonk},
  {Rafferty}, {Sabater}, {Smith}, {Schwarz}, {Smirnov}, {O'Sullivan},
  {Vedantham}, {White}, {Albert}, {Alegre}, {Asabere}, {Bacon}, {Bonafede},
  {Bonnassieux}, {Brienza}, {Bilicki}, {Bonato}, {Calistro Rivera}, {Cassano},
  {Cochrane}, {Croston}, {Cuciti}, {Dallacasa}, {Danezi}, {Dettmar}, {Di
  Gennaro}, {Edler}, {En{\ss}lin}, {Emig}, {Franzen}, {Garc{\'\i}a-Vergara},
  {Grange}, {G{\"u}rkan}, {Hajduk}, {Heald}, {Heesen}, {Hoang}, {Hoeft},
  {Horellou}, {Iacobelli}, {Jamrozy}, {Jeli{\'c}}, {Kondapally}, {Kukreti},
  {Kunert-Bajraszewska}, {Magliocchetti}, {Mahatma}, {Ma{\l}ek}, {Mandal},
  {Massaro}, {Meyer-Zhao}, {Mingo}, {Mostert}, {Nair}, {Nakoneczny},
  {Nikiel-Wroczy{\'n}ski}, {Orr{\'u}}, {Pajdosz-{\'S}mierciak}, {Pasini},
  {Prandoni}, {van Piggelen}, {Rajpurohit}, {Retana-Montenegro}, {Riseley},
  {Rowlinson}, {Saxena}, {Schrijvers}, {Sweijen}, {Siewert}, {Timmerman},
  {Vaccari}, {Vink}, {West}, {Wo{\l}owska}, {Zhang}, \&
  {Zheng}}]{2022A&A...659A...1S}
{Shimwell}, T.~W., {Hardcastle}, M.~J., {Tasse}, C., {et~al.} 2022, \aap, 659,
  A1

\bibitem[{{Shiohira} {et~al.}(2024){Shiohira}, {Fujii}, {Kita}, {Kimura},
  {Terada}, \& {Takahashi}}]{2024MNRAS.528.2136S}
{Shiohira}, Y., {Fujii}, Y., {Kita}, H., {et~al.} 2024, \mnras, 528, 2136

\bibitem[{{Siewert} {et~al.}(2021){Siewert}, {Schmidt-Rubart}, \&
  {Schwarz}}]{2021A&A...653A...9S}
{Siewert}, T.~M., {Schmidt-Rubart}, M., \& {Schwarz}, D.~J. 2021, \aap, 653, A9

\bibitem[{{Skrutskie} {et~al.}(2006){Skrutskie}, {Cutri}, {Stiening},
  {Weinberg}, {Schneider}, {Carpenter}, {Beichman}, {Capps}, {Chester},
  {Elias}, {Huchra}, {Liebert}, {Lonsdale}, {Monet}, {Price}, {Seitzer},
  {Jarrett}, {Kirkpatrick}, {Gizis}, {Howard}, {Evans}, {Fowler}, {Fullmer},
  {Hurt}, {Light}, {Kopan}, {Marsh}, {McCallon}, {Tam}, {Van Dyk}, \&
  {Wheelock}}]{2MASS}
{Skrutskie}, M.~F., {Cutri}, R.~M., {Stiening}, R., {et~al.} 2006, \aj, 131,
  1163

\bibitem[{{Slee}(1963)}]{1963Natur.199..991S}
{Slee}, O.~B. 1963, \nat, 199, 991

\bibitem[{{Stein} {et~al.}(2021){Stein}, {Vollmer}, {Boch}, {Landais},
  {Vannier}, {Brouty}, {Allen}, {Derriere}, \& {Ocvirk}}]{2021A&A...655A..17S}
{Stein}, Y., {Vollmer}, B., {Boch}, T., {et~al.} 2021, \aap, 655, A17

\bibitem[{{Suresh} {et~al.}(2020){Suresh}, {Chatterjee}, {Cordes}, {Bastian},
  \& {Hallinan}}]{2020ApJ...904..138S}
{Suresh}, A., {Chatterjee}, S., {Cordes}, J.~M., {Bastian}, T.~S., \&
  {Hallinan}, G. 2020, \apj, 904, 138

\bibitem[{{Tandoi} {et~al.}(2024){Tandoi}, {Guns}, {Foster}, {Ade}, {Anderson},
  {Ansarinejad}, {Archipley}, {Balkenhol}, {Benabed}, {Bender}, {Benson},
  {Bianchini}, {Bleem}, {Bouchet}, {Bryant}, {Camphuis}, {Carlstrom}, {Cecil},
  {Chang}, {Chaubal}, {Chichura}, {Chou}, {Coerver}, {Crawford}, {Cukierman},
  {Daley}, {de Haan}, {Dibert}, {Dobbs}, {Doussot}, {Dutcher}, {Everett},
  {Feng}, {Ferguson}, {Fichman}, {Galli}, {Gambrel}, {Gardner}, {Ge},
  {Goeckner-Wald}, {Gualtieri}, {Guidi}, {Halverson}, {Hivon}, {Holder},
  {Holzapfel}, {Hood}, {Huang}, {K{\'e}ruzor{\'e}}, {Knox}, {Korman},
  {Kornoelje}, {Kuo}, {Lee}, {Levy}, {Lowitz}, {Lu}, {Maniyar}, {Menanteau},
  {Millea}, {Montgomery}, {Moon}, {Nakato}, {Natoli}, {Noble}, {Novosad},
  {Omori}, {Padin}, {Pan}, {Paschos}, {Phadke}, {Prabhu}, {Qu}, {Quan},
  {Rahimi}, {Rahlin}, {Reichardt}, {Reuter}, {Rouble}, {Ruhl}, {Schiappucci},
  {Smecher}, {Sobrin}, {Stark}, {Stephen}, {Suzuki}, {Thompson}, {Thorne},
  {Trendafilova}, {Tucker}, {Umilta}, {Vieira}, {Wan}, {Wang}, {Whitehorn},
  {Wu}, {Yefremenko}, {Young}, \& {Zebrowski}}]{2024arXiv240113525T}
{Tandoi}, C., {Guns}, S., {Foster}, A., {et~al.} 2024, arXiv e-prints,
  arXiv:2401.13525

\bibitem[{{Tingay} {et~al.}(2012){Tingay}, {Goeke}, {Hewitt}, {Morgan},
  {Remillard}, {Williams}, {Bowman}, {Emrich}, {Ord}, {Booler}, {Crosse},
  {Pallot}, {Arcus}, {Colegate}, {Hall}, {Herne}, {Lynch}, {Schlagenhaufer},
  {Tremblay}, {Wayth}, {Waterson}, {Mitchell}, {Sault}, {Webster}, {Wyithe},
  {Morales}, {Hazelton}, {Wicenec}, {Williams}, {Barnes}, {Bernardi},
  {Greenhill}, {Kasper}, {Briggs}, {McKinley}, {Bunton}, {deSouza}, {Koenig},
  {Pathikulangara}, {Stevens}, {Cappallo}, {Corey}, {Kincaid}, {Kratzenberg},
  {Lonsdale}, {McWhirter}, {Rogers}, {Salah}, {Whitney}, {Deshpande}, {Prabu},
  {Roshi}, {Udaya-Shankar}, {Srivani}, {Subrahmanyan}, {Gaensler},
  {Johnston-Hollitt}, {Kaplan}, \& {Oberoi}}]{2012rsri.confE..36T}
{Tingay}, S., {Goeke}, R., {Hewitt}, J.~N., {et~al.} 2012, in Resolving The Sky
  - Radio Interferometry: Past, Present and Future, 36

\bibitem[{{Toet} {et~al.}(2021){Toet}, {Vedantham}, {Callingham}, {Veken},
  {Shimwell}, {Zarka}, {R{\"o}ttgering}, \& {Drabent}}]{2021A&A...654A..21T}
{Toet}, S.~E.~B., {Vedantham}, H.~K., {Callingham}, J.~R., {et~al.} 2021, \aap,
  654, A21

\bibitem[{{Tovmassian} {et~al.}(2017){Tovmassian}, {Gonz{\'a}lez-Buitrago},
  {Thorstensen}, {Kotze}, {Breytenbach}, {Schwope}, {Bernardini}, {Zharikov},
  {Hernandez}, {Buckley}, {de Miguel}, {Hambsch}, {Myers}, {Goff}, {Cejudo},
  {Starkey}, {Campbell}, {Ulowetz}, {Stein}, {Nelson}, {Reichart}, {Haislip},
  {Ivarsen}, {LaCluyze}, {Moore}, \& {Miroshnichenko}}]{2017A&A...608A..36T}
{Tovmassian}, G., {Gonz{\'a}lez-Buitrago}, D., {Thorstensen}, J., {et~al.}
  2017, \aap, 608, A36

\bibitem[{{van Haarlem} {et~al.}(2013){van Haarlem}, {Wise}, {Gunst}, {Heald},
  {McKean}, {Hessels}, {de Bruyn}, {Nijboer}, {Swinbank}, {Fallows},
  {Brentjens}, {Nelles}, {Beck}, {Falcke}, {Fender}, {H{\"o}randel},
  {Koopmans}, {Mann}, {Miley}, {R{\"o}ttgering}, {Stappers}, {Wijers},
  {Zaroubi}, {van den Akker}, {Alexov}, {Anderson}, {Anderson}, {van Ardenne},
  {Arts}, {Asgekar}, {Avruch}, {Batejat}, {B{\"a}hren}, {Bell}, {Bell}, {van
  Bemmel}, {Bennema}, {Bentum}, {Bernardi}, {Best}, {B{\^\i}rzan}, {Bonafede},
  {Boonstra}, {Braun}, {Bregman}, {Breitling}, {van de Brink}, {Broderick},
  {Broekema}, {Brouw}, {Br{\"u}ggen}, {Butcher}, {van Cappellen}, {Ciardi},
  {Coenen}, {Conway}, {Coolen}, {Corstanje}, {Damstra}, {Davies}, {Deller},
  {Dettmar}, {van Diepen}, {Dijkstra}, {Donker}, {Doorduin}, {Dromer}, {Drost},
  {van Duin}, {Eisl{\"o}ffel}, {van Enst}, {Ferrari}, {Frieswijk}, {Gankema},
  {Garrett}, {de Gasperin}, {Gerbers}, {de Geus}, {Grie{\ss}meier}, {Grit},
  {Gruppen}, {Hamaker}, {Hassall}, {Hoeft}, {Holties}, {Horneffer}, {van der
  Horst}, {van Houwelingen}, {Huijgen}, {Iacobelli}, {Intema}, {Jackson},
  {Jelic}, {de Jong}, {Juette}, {Kant}, {Karastergiou}, {Koers}, {Kollen},
  {Kondratiev}, {Kooistra}, {Koopman}, {Koster}, {Kuniyoshi}, {Kramer},
  {Kuper}, {Lambropoulos}, {Law}, {van Leeuwen}, {Lemaitre}, {Loose}, {Maat},
  {Macario}, {Markoff}, {Masters}, {McFadden}, {McKay-Bukowski}, {Meijering},
  {Meulman}, {Mevius}, {Middelberg}, {Millenaar}, {Miller-Jones}, {Mohan},
  {Mol}, {Morawietz}, {Morganti}, {Mulcahy}, {Mulder}, {Munk}, {Nieuwenhuis},
  {van Nieuwpoort}, {Noordam}, {Norden}, {Noutsos}, {Offringa}, {Olofsson},
  {Omar}, {Orr{\'u}}, {Overeem}, {Paas}, {Pandey-Pommier}, {Pandey}, {Pizzo},
  {Polatidis}, {Rafferty}, {Rawlings}, {Reich}, {de Reijer}, {Reitsma},
  {Renting}, {Riemers}, {Rol}, {Romein}, {Roosjen}, {Ruiter}, {Scaife}, {van
  der Schaaf}, {Scheers}, {Schellart}, {Schoenmakers}, {Schoonderbeek},
  {Serylak}, {Shulevski}, {Sluman}, {Smirnov}, {Sobey}, {Spreeuw}, {Steinmetz},
  {Sterks}, {Stiepel}, {Stuurwold}, {Tagger}, {Tang}, {Tasse}, {Thomas},
  {Thoudam}, {Toribio}, {van der Tol}, {Usov}, {van Veelen}, {van der Veen},
  {ter Veen}, {Verbiest}, {Vermeulen}, {Vermaas}, {Vocks}, {Vogt}, {de Vos},
  {van der Wal}, {van Weeren}, {Weggemans}, {Weltevrede}, {White}, {Wijnholds},
  {Wilhelmsson}, {Wucknitz}, {Yatawatta}, {Zarka}, {Zensus}, \& {van
  Zwieten}}]{2013A&A...556A...2V}
{van Haarlem}, M.~P., {Wise}, M.~W., {Gunst}, A.~W., {et~al.} 2013, \aap, 556,
  doi:\url{10.1051/0004-6361/201220873}

\bibitem[{{Vedantham} {et~al.}(2022){Vedantham}, {Callingham}, {Shimwell},
  {Benz}, {Hajduk}, {Ray}, {Tasse}, \& {Drabent}}]{2022ApJ...926L..30V}
{Vedantham}, H.~K., {Callingham}, J.~R., {Shimwell}, T.~W., {et~al.} 2022,
  \apjl, 926, L30

\bibitem[{{Vedantham} {et~al.}(2020){Vedantham}, {Callingham}, {Shimwell},
  {Tasse}, {Pope}, {Bedell}, {Snellen}, {Best}, {Hardcastle}, {Haverkorn},
  {Mechev}, {O'Sullivan}, {R{\"o}ttgering}, \& {White}}]{2020NatAs...4..577V}
---. 2020, Nature Astronomy, 4, 577

\bibitem[{{Villadsen} \& {Hallinan}(2019)}]{2019ApJ...871..214V}
{Villadsen}, J., \& {Hallinan}, G. 2019, \apj, 871, 214

\bibitem[{{Voges} {et~al.}(1999){Voges}, {Aschenbach}, {Boller},
  {Br{\"a}uninger}, {Briel}, {Burkert}, {Dennerl}, {Englhauser}, {Gruber},
  {Haberl}, {Hartner}, {Hasinger}, {K{\"u}rster}, {Pfeffermann}, {Pietsch},
  {Predehl}, {Rosso}, {Schmitt}, {Tr{\"u}mper}, \& {Zimmermann}}]{ROSAT_RASS}
{Voges}, W., {Aschenbach}, B., {Boller}, T., {et~al.} 1999, \aap, 349, 389

\bibitem[{{Wagenveld} {et~al.}(2023){Wagenveld}, {Kl{\"o}ckner}, \&
  {Schwarz}}]{2023A&A...675A..72W}
{Wagenveld}, J.~D., {Kl{\"o}ckner}, H.~R., \& {Schwarz}, D.~J. 2023, \aap, 675,
  A72

\bibitem[{{Walter} {et~al.}(1997){Walter}, {Hering}, \& {de
  Vegt}}]{1997A&AS..122..529W}
{Walter}, H.~G., {Hering}, R., \& {de Vegt}, C. 1997, \aaps, 122, 529

\bibitem[{{Wang} {et~al.}(2023){Wang}, {Murphy}, {Lenc}, {Mercorelli},
  {Driessen}, {Pritchard}, {Lao}, {Kaplan}, {An}, {Bannister}, {Heald}, {Lu},
  {Tuntsov}, {Walker}, \& {Zic}}]{Yuanming_ShortTimescale}
{Wang}, Y., {Murphy}, T., {Lenc}, E., {et~al.} 2023, \mnras, 523, 5661

\bibitem[{{Wendker}(1978)}]{Wendker_1978}
{Wendker}, H.~J. 1978, Astronomische Abhandlungen der Hamburger Sternwarte, 10,
  3

\bibitem[{{Wendker}(1987)}]{Wendker_1987}
---. 1987, \aaps, 69, 87

\bibitem[{{Wendker}(1995)}]{Wendker_1995}
---. 1995, \aaps, 109, 177

\bibitem[{{Wenger} {et~al.}(2000{\natexlab{a}}){Wenger}, {Ochsenbein}, {Egret},
  {Dubois}, {Bonnarel}, {Borde}, {Genova}, {Jasniewicz}, {Lalo{\"e}},
  {Lesteven}, \& {Monier}}]{simbad}
{Wenger}, M., {Ochsenbein}, F., {Egret}, D., {et~al.} 2000{\natexlab{a}},
  \aaps, 143, 9

\bibitem[{{Wenger} {et~al.}(2000{\natexlab{b}}){Wenger}, {Ochsenbein}, {Egret},
  {Dubois}, {Bonnarel}, {Borde}, {Genova}, {Jasniewicz}, {Lalo{\"e}},
  {Lesteven}, \& {Monier}}]{2000A&AS..143....9W}
---. 2000{\natexlab{b}}, Astronomy and Astrophysics Supplement Series, 143, 9

\bibitem[{{Whitehorn} {et~al.}(2016){Whitehorn}, {Natoli}, {Ade}, {Austermann},
  {Beall}, {Bender}, {Benson}, {Bleem}, {Carlstrom}, {Chang}, {Chiang}, {Cho},
  {Citron}, {Crawford}, {Crites}, {de Haan}, {Dobbs}, {Everett}, {Gallicchio},
  {George}, {Gilbert}, {Halverson}, {Harrington}, {Henning}, {Hilton},
  {Holder}, {Holzapfel}, {Hoover}, {Hou}, {Hrubes}, {Huang}, {Hubmayr},
  {Irwin}, {Keisler}, {Knox}, {Lee}, {Leitch}, {Li}, {McMahon}, {Meyer},
  {Mocanu}, {Nibarger}, {Novosad}, {Padin}, {Pryke}, {Reichardt}, {Ruhl},
  {Saliwanchik}, {Sayre}, {Schaffer}, {Smecher}, {Stark}, {Story}, {Tucker},
  {Vanderlinde}, {Vieira}, {Wang}, \& {Yefremenko}}]{2016ApJ...830..143W}
{Whitehorn}, N., {Natoli}, T., {Ade}, P.~A.~R., {et~al.} 2016, \apj, 830, 143

\bibitem[{{Whiting} \& {Humphreys}(2012)}]{2012PASA...29..371W}
{Whiting}, M., \& {Humphreys}, B. 2012, \pasa, 29, 371

\bibitem[{{Whiting}(2012)}]{2012MNRAS.421.3242W}
{Whiting}, M.~T. 2012, \mnras, 421, 3242

\bibitem[{{Williams} \& {Berger}(2015)}]{2015ApJ...808..189W}
{Williams}, P.~K.~G., \& {Berger}, E. 2015, \apj, 808, 189

\bibitem[{{Williams} {et~al.}(2014){Williams}, {Cook}, \&
  {Berger}}]{2014ApJ...785....9W}
{Williams}, P.~K.~G., {Cook}, B.~A., \& {Berger}, E. 2014, \apj, 785, 9

\bibitem[{{Wright} {et~al.}(2010){Wright}, {Eisenhardt}, {Mainzer}, {Ressler},
  {Cutri}, {Jarrett}, {Kirkpatrick}, {Padgett}, {McMillan}, {Skrutskie},
  {Stanford}, {Cohen}, {Walker}, {Mather}, {Leisawitz}, {Gautier}, {McLean},
  {Benford}, {Lonsdale}, {Blain}, {Mendez}, {Irace}, {Duval}, {Liu}, {Royer},
  {Heinrichsen}, {Howard}, {Shannon}, {Kendall}, {Walsh}, {Larsen}, {Cardon},
  {Schick}, {Schwalm}, {Abid}, {Fabinsky}, {Naes}, \&
  {Tsai}}]{2010AJ....140.1868W}
{Wright}, E.~L., {Eisenhardt}, P. R.~M., {Mainzer}, A.~K., {et~al.} 2010, \aj,
  140, 1868

\bibitem[{{Yiu} {et~al.}(2024){Yiu}, {Vedantham}, {Callingham}, \&
  {G{\"u}nther}}]{2023arXiv231207162Y}
{Yiu}, T.~W.~H., {Vedantham}, H.~K., {Callingham}, J.~R., \& {G{\"u}nther},
  M.~N. 2024, \aap, 684, A3

\bibitem[{{Zacharias} {et~al.}(2013){Zacharias}, {Finch}, {Girard}, {Henden},
  {Bartlett}, {Monet}, \& {Zacharias}}]{UCAC4}
{Zacharias}, N., {Finch}, C.~T., {Girard}, T.~M., {et~al.} 2013, \aj, 145, 44

\end{thebibliography}

\appendix

\section{\stab\ and \rtab\ excerpts}
\label{app: table excerpts}

\clearpage
\onecolumn

\begin{landscape}
\centering
\begin{small}
\begin{longtable}{rrrrrrrr}
\caption[]{\label{tab: stars table demo}{The first ten rows of the Stars.dat table. As the Stars.dat table has many columns, we have split the columns and added a ``row'' column to include each continuous row.
}}\\
Row &Identifier & Simbad & Gaia & Tycho & 2MASS & GCVS \\
\hline
1 & SRSC 00000 & CPD-62  4126 & Gaia DR3 5853502492753436416 & TYC 9010-2749-1 & 2MASS J14270430-6246553 &  \\
2 & SRSC 00001 & SCR J2241-6119A & Gaia DR3 6406967509044836096 &  & 2MASS J22414436-6119311 &  \\
3 & SRSC 00002 & CD-30  6530 & Gaia DR3 5640708647248169600 & TYC 7136-2264-1 & 2MASS J08355977-3042306 &  \\
4 & SRSC 00003 & CD-49   451 & Gaia DR3 4917782741272766592 & TYC 8043-814-1 & 2MASS J01372081-4911443 &  \\
5 & SRSC 00004 & ATO J183.5330+47.2673 & Gaia DR3 1545247048904161664 &  & 2MASS J12140814+4716038 &  \\
6 & SRSC 00005 & 2MASS J11414215-6521298 & Gaia DR3 5236539580341818624 &  & 2MASS J11414215-6521298 &  \\
7 & SRSC 00006 & Gaia DR3 6090676737160646528 & Gaia DR3 6090676737160646528 &  &  &  \\
8 & SRSC 00007 & UCAC4 053-023945 & Gaia DR3 6349781084650820608 &  & 2MASS J21072847-7926274 &  \\
9 & SRSC 00008 & PM J09551-0819 & Gaia DR3 3771533721761978496 & TYC 5475-507-1 & 2MASS J09550963-0819259 &  \\
10 & SRSC 00009 & Gaia DR3 5971238129102004992 & Gaia DR3 5971238129102004992 &  &  &  \\
\hline
 &  &  &  &  &  & \\
Row &HIP & UCAC4 & Survey & Survey\_id & Epoch & RAdeg \\
\hline
1 &  & UCAC4 137-115021 & GaiaDR3 & Gaia DR3 5853502492753436416 & J2016 & 216.767919155244 \\
2 &  & UCAC4 144-211738 & GaiaDR3 & Gaia DR3 6406967509044836096 & J2016 & 340.436272395047 \\
3 &  & UCAC4 297-048519 & GaiaDR3 & Gaia DR3 5640708647248169600 & J2016 & 128.998712548091 \\
4 & HIP 7554 & UCAC4 205-001600 & GaiaDR3 & Gaia DR3 4917782741272766592 & J2016 & 24.34012397651 \\
5 &  & UCAC4 687-053931 & GaiaDR3 & Gaia DR3 1545247048904161664 & J2016 & 183.532979271787 \\
6 &  & UCAC4 124-048667 & GaiaDR3 & Gaia DR3 5236539580341818624 & J2016 & 175.425221260026 \\
7 &  &  & GaiaDR3 & Gaia DR3 6090676737160646528 & J2016 & 211.790382522137 \\
8 &  & UCAC4 053-023945 & GaiaDR3 & Gaia DR3 6349781084650820608 & J2016 & 316.869374569023 \\
9 &  & UCAC4 409-049725 & GaiaDR3 & Gaia DR3 3771533721761978496 & J2016 & 148.789425163571 \\
10 &  &  & GaiaDR3 & Gaia DR3 5971238129102004992 & J2016 & 253.074412168032 \\
\hline
 &  &  &  &  &  & \\
Row &e\_RAdeg & DEdeg & e\_DEdeg & plx & e\_plx & pmRA  \\
\hline
1 & 6e-06 & -62.782098179545 & 9e-06 & 1.53 & 0.01 & -3.841  \\
2 & 8e-06 & -61.325706059717 & 1e-05 & 35.2 & 0.01 & 150.4  \\
3 & 8e-06 & -30.70858670455 & 1e-05 & 15.7 & 0.01 & -64.11  \\
4 & 1e-05 & -49.195143947004 & 1e-05 & 45.12 & 0.01 & 496.1  \\
5 & 1e-05 & 47.267364915243 & 1e-05 & 25.76 & 0.02 & -137.2  \\
6 & 1e-05 & -65.358320600806 & 1e-05 & 9.92 & 0.02 & -40.18  \\
7 & 1e-05 & -50.034311567705 & 1e-05 & 10.99 & 0.02 & -54.59  \\
8 & 1e-05 & -79.441225054251 & 1e-05 & 14.44 & 0.02 & 28.66  \\
9 & 1e-05 & -8.323892037812 & 1e-05 & 31.53 & 0.02 & -135.68  \\
10 & 2e-05 & -38.189228568212 & 1e-05 & 6.08 & 0.02 & -11.18  \\
\hline
 &  &  &  &  &  & \\
Row &e\_pmRA & pmDE & e\_pmDE & Radio\_multiple & Identification\_method  &  \\
\hline
1 & 0.007 & -5.44 & 0.01 & 0.0 & 18  &  \\
2 & 0.01 & -87.89 & 0.01 & 0.0 & 18  &  \\
3 & 0.01 & -14.3 & 0.01 & 0.0 & 18  &  \\
4 & 0.01 & 116.1 & 0.01 & 0.0 & 18  &  \\
5 & 0.01 & -72.58 & 0.01 & 0.0 & 16  &  \\
6 & 0.02 & -4.38 & 0.02 & 0.0 & 18  &  \\
7 & 0.02 & -46.85 & 0.02 & 0.0 & 16  &  \\
8 & 0.02 & -61.04 & 0.02 & 0.0 & 16  &  \\
9 & 0.02 & -7.06 & 0.02 & 0.0 & 18  &  \\
10 & 0.02 & -22.68 & 0.02 & 0.0 & 16  &  \\
\hline 
    % \end{tabular}    
%  \end{table}
\end{longtable}
\end{small}
\end{landscape} 

\clearpage
\onecolumn

\begin{landscape}
\centering
\begin{small}
\begin{longtable}{rrrrrrrrrrrr}
\caption[]{\label{tab: radio table demo}{The first ten rows of the Radio.dat table. As the Radio.dat table has many columns, we have split the columns and added a ``row'' column to include each continuous row.
}}\\
Row &Identifier & Match\_separation & Radio\_id & Field & Obs\_date & Exposure & RAdeg & DEdeg & e\_RAdeg & e\_DEdeg \\
\hline
1 & SRSC 00000 & 1.2 & SB50301\_component\_4182a & 50301.0 & 2023-06-02T12:53:55 & 716.0 & 216.7685 & -62.7823 & 2.0 & 2.0 \\
2 & SRSC 00000 &  & SB47240\_component\_3629a & 47240.0 & 2023-01-20T21:45:43 & 726.0 & 216.7677 & -62.783 &  &  \\
3 & SRSC 00000 & 1.7 & SB51817\_component\_4823a & 51817.0 & 2023-08-03T08:56:08 & 727.0 & 216.7675 & -62.7817 & 2.0 & 2.0 \\
4 & SRSC 00000 & 1.6 & SB45665\_component\_4457a & 45665.0 & 2022-11-19T01:13:36 & 717.0 & 216.767 & -62.7823 & 2.0 & 2.0 \\
5 & SRSC 00000 & 0.7 & SB45667\_component\_4705a & 45667.0 & 2022-11-19T01:40:28 & 717.0 & 216.7676 & -62.7823 & 2.0 & 2.0 \\
6 & SRSC 00000 & 0.7 & SB45821\_component\_4152b & 45821.0 & 2022-11-24T23:41:57 & 36011.0 & 216.7681 & -62.7819 & 2.0 & 2.0 \\
7 & SRSC 00000 & 2.5 & SB51522\_component\_4617a & 51522.0 & 2023-07-21T10:00:19 & 726.0 & 216.7664 & -62.782 & 2.0 & 2.0 \\
8 & SRSC 00000 & 2.1 & SB53482\_component\_4650a & 53482.0 & 2023-10-04T03:45:51 & 737.0 & 216.7689 & -62.7817 & 2.0 & 2.0 \\
9 & SRSC 00000 & 1.1 & SB22350\_component\_2094a & 22350.0 & 2021-02-01T20:56:05 & 895.0 & 216.7673 & -62.7821 & 2.0 & 2.0 \\
10 & SRSC 00000 & 2.2 & SB47236\_component\_4433a & 47236.0 & 2023-01-20T20:45:20 & 726.0 & 216.7684 & -62.7827 & 2.0 & 2.0 \\
\hline
 &  &  &  &  &  &  &  &  &  & \\
Row &Freq & SpeakI & e\_SpeakI & StotI & e\_StotI & bmax & bmin & PA & e\_bmax & e\_bmin \\
\hline
1 & 887.5 & 1.6 & 0.3 & 1.5 & 0.5 & 17.0 & 11.0 & 123.01 & 3.0 & 2.0 \\
2 & 887.5 & 2.1 & 0.2 & 2.8 & 0.4 &  &  &  &  &  \\
3 & 887.5 & 1.2 & 0.2 & 1.5 & 0.3 & 18.0 & 13.0 & 22.94 & 2.0 & 2.0 \\
4 & 887.5 & 1.6 & 0.2 & 1.8 & 0.3 & 19.0 & 11.0 & 134.44 & 2.0 & 1.0 \\
5 & 887.5 & 1.4 & 0.2 & 1.7 & 0.4 & 16.0 & 15.0 & 86.21 & 2.0 & 2.0 \\
6 & 943.5 & 0.64 & 0.03 & 0.74 & 0.05 & 17.5 & 15.0 & 133.57 & 0.7 & 0.7 \\
7 & 887.5 & 1.4 & 0.2 & 1.6 & 0.4 & 22.0 & 10.0 & 141.85 & 2.0 & 2.0 \\
8 & 887.5 & 1.3 & 0.2 & 1.4 & 0.4 & 17.0 & 13.0 & 133.59 & 3.0 & 2.0 \\
9 & 1367.5 & 1.6 & 0.3 & 1.6 & 0.5 & 9.0 & 8.0 & 170.91 & 2.0 & 1.0 \\
10 & 887.5 & 1.4 & 0.2 & 1.7 & 0.4 & 20.0 & 11.0 & 135.82 & 2.0 & 2.0 \\
\hline
 &  &  &  &  &  &  &  &  &  & \\
Row &Survey & SpeakV & e\_SpeakV & StotV & e\_StotV & Telescope & localrmsV & localrmsI & Ref  & \\
\hline
1 &  &  &  &  &  & ASKAP &  & 0.26 & ThisWork  & \\
2 & VAST 23e & 1.3 &  & 1.9 &  & ASKAP &  & 0.29 & 10.1093/mnras/stab299  & \\
3 &  &  &  &  &  & ASKAP &  & 0.22 & ThisWork  & \\
4 &  &  &  &  &  & ASKAP &  & 0.21 & ThisWork  & \\
5 &  &  &  &  &  & ASKAP &  & 0.24 & ThisWork  & \\
6 & 10-11hour &  &  &  &  & ASKAP &  & 0.03 & ThisWork  & \\
7 &  &  &  &  &  & ASKAP &  & 0.25 & ThisWork  & \\
8 &  &  &  &  &  & ASKAP &  & 0.24 & ThisWork  & \\
9 &  &  &  &  &  & ASKAP &  & 0.28 & ThisWork  & \\
10 &  &  &  &  &  & ASKAP &  & 0.24 & ThisWork  & \\
\hline 
    % \end{tabular}    
%  \end{table}
\end{longtable}
\end{small}
\end{landscape}

\end{document}